\documentclass[twocolumn]{aastex631}
\usepackage{amsmath}

\usepackage{hyperref}

\usepackage{enumitem}

\def\Wm {W~m$^{-2}$}
\def\cm {cm$^{-3}$}

\def\ic {IC\,10}
\def\hii {H{\scriptsize II}}
\def\density {$n_{\rm e}$}
\def \neoiii {$n_{\rm e\, [O\,\textsc{iii}]}$}
\def\nesiii {$n_{\rm e\, [S\,\textsc{iii}]}$}

\def\oiii {[O\,{\sc iii}]}
\def\siii {[S\,{\sc iii}]}
\def\oiiiup {[O\,{\sc iii}]\,$\lambda88\mu$m}
\def\oiiilw {[O\,{\sc iii}]\,$\lambda52\mu$m}
\def\siiiup {[S\,{\sc iii}]\,$\lambda33\mu$m}
\def\siiilw {[S\,{\sc iii}]\,$\lambda18\mu$m}
 
\def\oib {[O\,{\sc i}]\,$\lambda145.5\mu$m}

\def\ha {H{\sc $\alpha$}}

\begin{document}

\title{Electron density distribution in \hii\,regions in IC\,10}

\correspondingauthor{Fiorella L. Polles}
\email{fpolles@usra.edu}

\author[0000-0003-0347-3201]{Fiorella L. Polles}
\affiliation{SOFIA Science Center, USRA, NASA Ames Research Center, M.S. N232-12 Moffett Field, CA, 94035, USA}

\author[0000-0002-3698-7076]{Dario Fadda}
\affiliation{Space Telescope Science Institute, 3700 San Martin Drive, Baltimore, MD, 21218, USA}

\author[0000-0002-9123-0068]{William D. Vacca}
\affiliation{Gemini Observatory/NSF's NOIRLab, 950 N. Cherry Ave, Tucson, AZ, 85719, USA}

\author[0000-0003-1791-723X]{Nicholas P. Abel}
\affiliation{University of Cincinnati, Clermont College, 4200 Clermont College Drive, Batavia, OH, 45103, USA}

\author[0000-0002-5635-5180]{Mélanie Chevance}
\affiliation{Universit\"{a}t Heidelberg, Zentrum f\"{u}r Astronomie, Institut f\"{u}r Theoretische Astrophysik, Albert-Ueberle-Str 2, D-69120 Heidelberg, Germany}
\affiliation{Cosmic Origins Of Life (COOL) Research DAO, \href{https://coolresearch.io}{coolresearch.io}}

\author[0000-0002-7299-8661]{Christian Fischer}
\affiliation{Deutsche SOFIA Institut, University of Stuttgart, D-70569 Stuttgart, Germany}

\author[0000-0002-3466-6164]{James M. Jackson}
\affiliation{Green Bank Observatory, P.O. Box 2, Green Bank, WV, 24944, USA}

\author[0000-0002-7716-6223]{Vianney Lebouteiller}
\affiliation{Université Paris-Saclay, Université Paris-Cité, CEA, CNRS, AIM, 91191, Gif-sur-Yvette, France}

\author[0000-0003-3229-2899]{Suzanne Madden}
\affiliation{Université Paris Cité, Université Paris-Saclay, CEA, CNRS, AIM, 91191, Gif-sur-Yvette, France}

\author[0000-0002-9190-9986]{Lise Ramambason}
\affiliation{Universit\"{a}t Heidelberg, Zentrum f\"{u}r Astronomie, Institut f\"{u}r Theoretische Astrophysik, Albert-Ueberle-Str 2, D-69120 Heidelberg, Germany}



\begin{abstract}
We present the \oiiilw\, map of the dwarf galaxy \ic\, obtained with the Field-Imaging Far-Infrared Line Spectrometer (FIFI-LS) on board the Stratospheric Observatory for Infrared Astronomy (SOFIA). We combine the \oiiilw\, map with {\it Herschel} and {\it Spitzer} observations, to estimate the electron density distribution of the brightest \hii\, regions of \ic. We find that the line ratio \oiiiup/\oiiilw\ gives electron density (\density) values (\neoiii) that cover a broad range, while the \density\ values obtained using the line ratio \siiiup/\siiilw\ (\nesiii) are all similar within the uncertainties. \neoiii\ is similar to \nesiii\ for the M1, M2 and A1 regions, and it is higher than \nesiii\ for the two regions, A2 and M1b, which are the brightest in the 24\,$\mu$m continuum emission. These results suggest that for these regions the two ions, O$^{++}$ and S$^{++}$, trace two different ionised gas components, and that the properties of the ionised gas component traced by the O$^{++}$ ion are more sensitive to the local physical conditions. In fact, while the gas layer traced by \siii\ does not keep track of the characteristics of the radiation field, the \neoiii\, correlates with the star formation rate (SFR), the dust temperature and the 24\,$\mu$m. Therefore, \neoiii\ is an indicator of the evolutionary stage of the \hii\, region and the radiation field, with higher \neoiii\, found in younger SF regions and in more energetic environments.

\end{abstract}

\keywords{Interstellar Medium (847) --- Dwarf galaxies (416)}


\section{Introduction} \label{sec:intro}
The gas density plays a role in many properties of the interstellar medium (ISM), such as the gas-pressure, the electron temperature, the ionization parameter and the elemental abundances (e.g., \citealt{rubin89}; \citealt{vaught23}; \citealt{mendez24}). The electron density (\density) is often approximated to be constant inside the \hii\, regions, until the ionization front when detailed information about the density is lacking \cite[e.g.,][]{cosens22, spinoglio22}. However, several studies have revealed that this assumption is incorrect: \density\, can have a complex radial profile inside the \hii\, regions. Some \hii\ regions show density gradients (e.g., \citealt{rubin11}; \citealt{mcleod16}), others show density fluctuations (e.g., \citealt{mendez23}; \citealt{vaught23}). Moreover, integrated \hii\ regions may include a mix of ionized gas layers with different densities, unresolved high-density clumps surrounded by lower-density ionised gas as well as filaments with different densities due to stellar feedback (e.g., \citealt{odell17}). The mix of gas densities is even more complex on the scale of a galaxy, especially for unresolved galaxies where all of the components are mixed within a spatial resolution element. The uncertainties in constraining \density\, propagate to uncertainties on the derived physical properties of the ISM, such as thermal gas pressure, and on the derived stellar feedback mechanisms. Thus, an accurate knowledge of \density\, is of fundamental importance to properly infer the structure and the properties of the ISM, as well as the interplay between ISM and stars.

The electron density can be directly derived using ratios of lines emitted by two levels of the same ion with different critical densities and energy of the transition, so that the excitation of these levels depends only on the density and the temperature of the gas \cite[e.g.,][]{osterbrock06, kewley19}. Examples of line ratios that can be used to derive the electron density are [Si\,{\sc iii}]\,$\lambda1883\AA$/Si\,{\sc iii}]\,$\lambda1892\AA$ and [C\,{\sc iii}]\,$\lambda1906\AA$/C\,{\sc iii}]\,$\lambda1909\AA$ in the UV \cite[e.g.,][]{nussbaumer79,keenan92}, [O\,{\sc ii}]$\lambda$ 3729 $\AA$/3727 $\AA$, [Ar\,{\sc iv}]$\lambda$ 4711 $\AA$/4740 $\AA$ and [S\,{\sc ii}]$\lambda$ 6716 $\AA$/6731 $\AA$ in the optical \cite[e.g.,][]{kaasinen17, dellabruna20}, and \oiiiup/52 $\mu$m, [Ne\,{\sc iii}]$\lambda$ 36 $\mu$m/15 $\mu$m, and [N\,{\sc ii}]$\lambda$ 205 $\mu$m/122 $\mu$m in the infrared (IR) \cite[e.g.,][]{pineda19,chevance20,peng21}. Based on the ionization potentials (IPs) and critical densities, the lines may arise from different components of \hii\ regions. Considering the \hii\ region as a single slab of gas illuminated by a stellar cluster, the ionization lines such as [Ne\,{\sc iii}]$\lambda$ 36 $\mu$m and [Ne\,{\sc iii}]$\lambda$ 15 $\mu$m (IP 40.96 eV), trace the ionised gas close to the ionizing source, while the lines with lower excitation potential, such as [N\,{\sc ii}]$\lambda$ 122 $\mu$m and [N\,{\sc ii}]$\lambda$ 205 $\mu$m (IP 14.53 eV), trace the ionization front \citep[e.g.,][]{kewley19}. Moreover, each electron density diagnostic is sensitive only to the density range below the critical densities of the lines of the ratio. 
Hence, we may expect to determine different density values for different line ratios, providing a convenient means to determine the density structure of the ionized gas \citep[e.g.,][]{rubin11, fernandez16}. The best density indicators are those for which the derived value depends only very weakly on the temperature, which is why low lying forbidden lines are the best (e.g., \citealt{spinoglio15}, \citealt{kewley19}).  Moreover, IR lines are less affected by extinction by dust than optical lines. Therefore, IR fine--structure line ratios emitted by the same ion are excellent density tracers. The IR wavelength domain provides several electron density diagnostics, each of them covering a different and complementary density range: e.g. [N\,{\sc ii}] traces a density range of $1 - 100$\,cm$^{-3}$, [O\,{\sc iii}] $10 - 10^3$\,cm$^{-3}$, [S\,{\sc iii}] a range between $100$ and $10^4$\,cm$^{-3}$, and the [Ar\,{\sc iii}] and [Ne\,{\sc iii}] lines are useful between $10^3$ and $10^5$\,cm$^{-3}$ (e.g., \citealt{osterbrock06}). 

Ideal laboratories to investigate the electron density distribution using IR diagnostics are nearby galaxies. Here we focus on the dwarf galaxy \ic. Despite the low metallicity of $12+\log({\rm O/H})=8.26$ ($\approx1/3$ solar; e.g., \citealt{lequeux79}; \citealt{magrini09}), this galaxy has a high surface density of Wolf-Rayet (WR) stars (e.g., \citealp{massey02}; \citealp{tehrani17}) and an unexpected high ratio of carbon-type WR (WC) to nitrogen-type WR (WN), indicating its starburst nature (e.g., \citealp{crowther03}). Due to these characteristics and its proximity ($\sim 715$\,kpc; e.g., \citealt{kim09}), \ic\ has been observed with several instruments, covering a broad range of wavelengths. The rich dataset extends from X-ray to radio, allowing the characterization of the stellar population \cite[e.g.,][]{crowther03, vacca07}, the physical and kinematic properties \cite[e.g.,][]{lopez11, cosens22} of the several \hii\ regions \cite[]{hodge91} as well as the neutral \cite[e.g.,][]{ashley14} and molecular gas  \cite[e.g.,][]{kepley18} of the galaxy. The properties of the ionised gas of \ic\ have been investigated in a previous paper (\citealt{polles19}), through models of the mid-- and far--infrared {\it Spitzer} and {\it Herschel} fine structure lines. That study revealed that most of the \hii\ regions have almost-identical physical properties (density, ionization parameter and age of the stellar cluster) and they are all matter-bounded. However, the analysis of the physical properties of the \hii\ regions was limited by the availability of a single electron density tracer, \siiiup/\siiilw. The lack of a variety of density tracers may have prevented us from unveiling the density structure of the \hii\ regions, 
which could be used to constrain and quantify the stellar feedback processes, and reveal the evolution of \hii\ regions. Assuming a single density could lead to incorrect conclusions regarding, for example, the determination of the elemental abundances.
In this paper, we present the new \oiiilw\ map of \ic\ observed with the Stratospheric Observatory for Infrared Astronomy (SOFIA; \citealt{Erickson93}), and we analyse the electron density structure of the brightest \hii\ regions of the galaxy, combining the SOFIA data with {\it Herschel} and {\it Spitzer} data.

The structure of the paper is the following: in Section~\ref{sec:data} we describe the data and in Section~\ref{sec:ne} we estimate the electron density distribution using the line ratios \oiiiup/\oiiilw\, and \siiiup/\siiilw. The results and their implications are discussed in Section~\ref{sec:discussion}. Finally, Section~\ref{sec:conclusion} summarizes the main points of this study. 

\section{Data}\label{sec:data}
The characteristics of the observations used in this study are presented in Table~\ref{tab:observations}, and the maps are shown in Figure~\ref{fig:maps}. The general properties of the IR fine--structure lines used to calculate \density\, are summarized in Table~\ref{tab:line_prop}.

\subsection{SOFIA/FIFI-LS data}
\subsubsection{Observations}
\ic\ has been observed with the Far Infrared Field--Imaging Line Spectrometer (FIFI-LS; \citealt{fischer18}; \citealt{colditz18}) on board SOFIA as part of the GTO program 70\_0908 (P.I. Alfred Krabbe), during two flights, August 31 and September 1, 2022, flying out of Palmdale, California.

FIFI-LS is an integral field far-infrared spectrometer, providing a spectrum at each pixel in its field of view (FOV), with a spectral resolution from $500$ to $2000$ from $51$\,$\mu$m to $203$\,$\mu$m. This instrument includes two channels for simultaneous observations: the blue channel covering the wavelength range $51 - 120$\,$\mu$m, and the red channel observing between $115 - 203$\,$\mu$m. Each channel has a $5\times5$ pixel projection on the sky. For the blue channel, the pixel size is $6\arcsec\times6\arcsec$, for the red channel it is $12\arcsec\times12\arcsec$. For each spatial pixel (also called ``spaxel'') the light is dispersed spectrally over 16 pixels, providing an integral-field data cube for each observation, covering a total spectral bandwidth between $1500$ and $3000$\,km\,s$^{-1}$, depending on the wavelength. The light is dispersed by ruled gratings mounted to a tilting drive and re-imaged onto a detector of stressed Gallium doped Germanium photo conductors.

The two main star--forming regions of \ic\, (see Figure~\ref{fig:maps}) have been covered in \oiiilw\, with the blue channel, and \oib\, with the red channel. 
In this paper we focus on the \oiiilw\, observations. 
The data was acquired in symmetric chop mode, i.e. the sky background positions are located across the optical axis from source position (see \href{https://irsa.ipac.caltech.edu/data/SOFIA/docs/instruments/handbooks/FIFI-LS_Handbook_for_Archive_Users_Ver1.0.pdf}{FIFI-LS Handbook for Archive Users}, Sec.2.4.1), for maximum observing efficiency \citep{Fischer16}. A chop throw of $4\arcmin$ towards the north--west as well as south--east direction was chosen to safely chop out of the galaxy. The total wall-clock time of the \oiiilw\, was about $5.5$\,h. Four blue channel $30\arcsec\times30\arcsec$ fields where pointed to regions with bright \oiiiup\ emission previously identified with {\it Herschel}/PACS (see Sec~\ref{sec:scpitzerherschel}) to create the map shown in Figure~\ref{fig:maps}. Additional sub-pixel dithering was performed to improve spatial sampling. The field on the bright spot in the north-western sub-map was integrated for a total time of about $1$\,h while the other fields were integrated for about $1.5$\,h each. The spectral resolution at \oiiilw\, is R\,=\,791 and the point spread function has a size of $6.6\arcsec$ (\citealt{fadda2023}; corresponding to $\approx23.1$\,pc) assuming a diffraction limited telescope.

\subsubsection{Data reduction, atmospheric correction and calibration}
\begin{figure*}[t!]
\centering
         \includegraphics[width=\textwidth, clip]{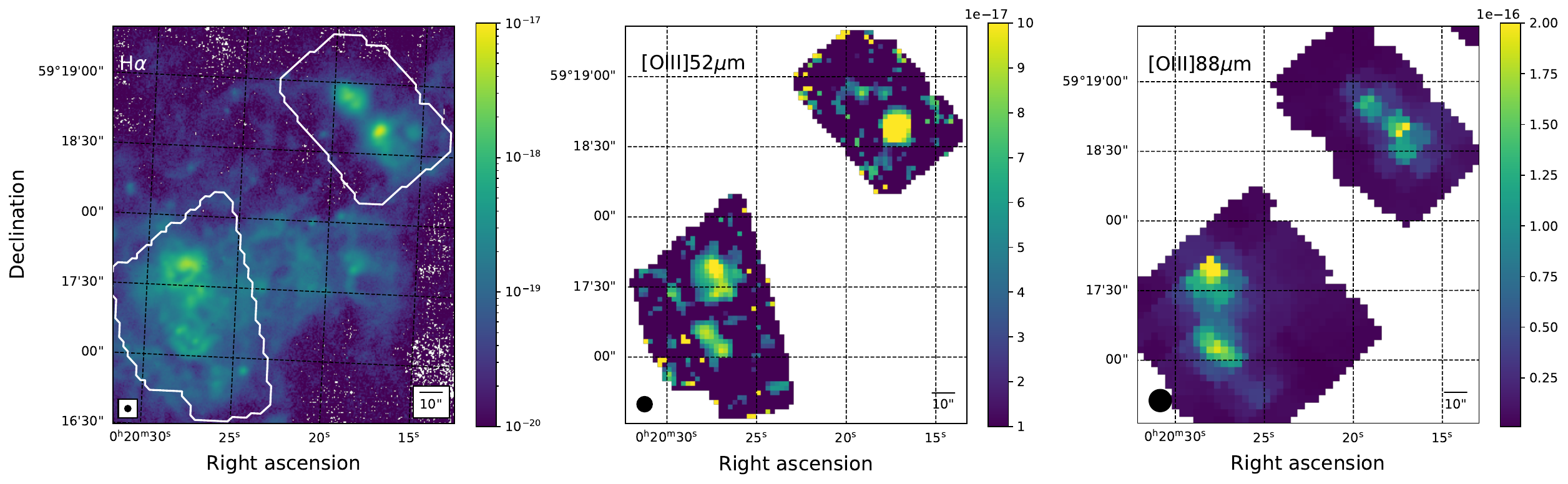}\\
         \includegraphics[width=\textwidth,clip]{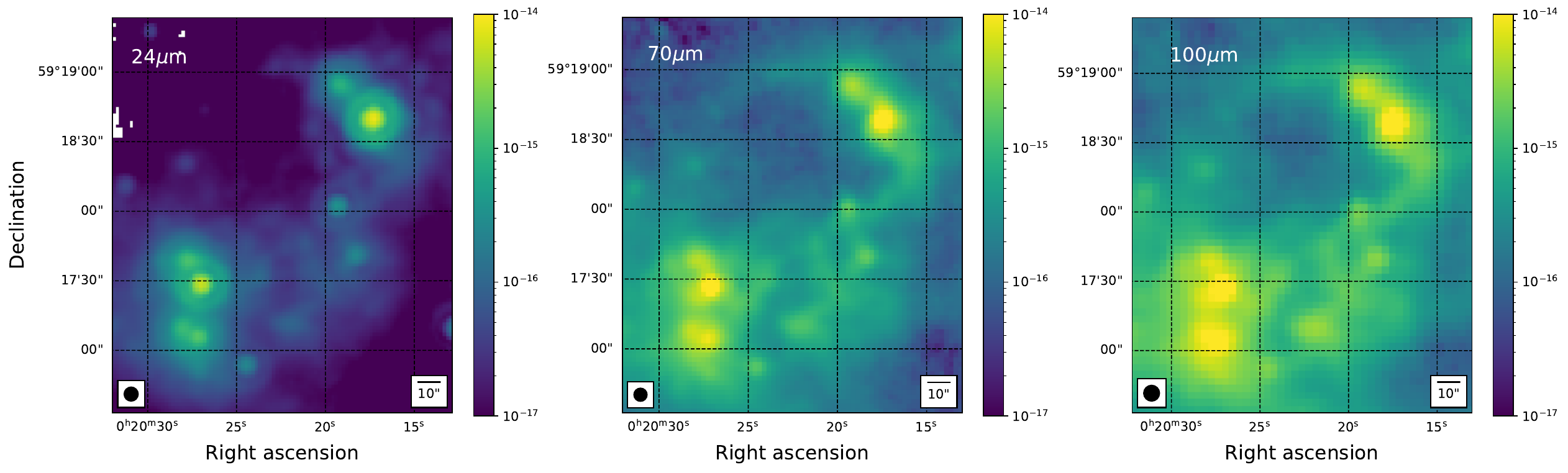}
    \caption{Intensity maps ($W\,m^{-2}pixel^{-1}$) of IC\,10. Top: \ha\, map with the regions observed in \oiiilw\, delineated in white (left), \oiiilw\, map (middle), \oiiiup\, map (right). Bottom: 24\,$\mu$m map (left), 70\,$\mu$m map (middle) and 100\,$\mu$m map (right). The black circles in the lower-left corner of each map correspond to the beam of each observation. The spatial resolution is about $2.2\arcsec$ for \ha, $6.6\arcsec$ for \oiiilw, $9.5\arcsec$ for \oiiiup, $6.0\arcsec$ for 24$\mu$m, $5.6\arcsec$ for 70$\mu$m and $6.8\arcsec$ for 100$\mu$m.}
    \label{fig:maps}
\end{figure*}

The data has been reduced with the FIFI-LS data reduction pipeline \citep{vacca20} using the calibration products described in \citet{fadda2023}. We applied a kernel size of 0.5$\times$FWHM and a window size of 1.5$\times$FWHM, which ensures that the spatial resolution is conserved. The output cube is sampled at $2\arcsec$ spatially and $40$\,km\,s$^{-1}$ spectrally. Atmospheric absorption may have a significant impact on the observed flux, even at the high altitudes of the flights, between $39\,000$ and $45\,000$ feet. To correct the data for the atmospheric absorption, the pipeline creates a transmission spectrum for the local conditions (altitude, elevation and water vapor overburden) for the observed nod cycle. Since the water vapor overburden was not continuously observed, we used satellite-derived values rescaled to direct measurements obtained during each flight between different observations \citep{iserlohe21,fischer21}.
The reduced cube has been corrected by the atmospheric absorption factor estimated at the wavelength corresponding to the position of \oiiilw\, using SOSPEX\footnote{\url{https://github.com/darioflute/sospex}} \citep{fadda18, sospex}, avoiding overcorrection of the wing of the \oiiilw\ line which falls into a deep telluric line. The absolute amplitude calibration uncertainty is assumed to be $15\%$, of which $10\%$ is relative uncertainty from FIFI-LS seen between flight series and different calibrators. The rest arises from atmospheric corrections and calibration modeling uncertainties. The total intensity map of \oiiilw\, is shown on Figure~\ref{fig:maps} for illustrative purposes only. We estimated the integrated fluxes of \oiiilw\ and \oiiiup\ directly from the cube (see Sect.~\ref{sec:neoiii}).

\subsection{Ancillary data}

\subsubsection{MIR and FIR data}\label{sec:scpitzerherschel}

\ic\, has been mapped in the infrared also by {\it Spitzer} and by {\it Herschel} (Dwarf Galaxy Survey (DGS); \citealt{madden13}) covering the mid-- (MIR) and far--infrared (FIR), respectively.  The MIR {\it Spitzer}/Infrared Spectrograph (IRS; \citealt{houck04}) and FIR {\it Herschel}/Photodetector Array Camera and Spectrometer (PACS; \citealt{poglitsch10}) spectroscopy data has been presented in \cite{polles19}. We are particularly interested in bringing into this study the two \siii\ maps observed with the {\it Spitzer}/IRS low-resolution (LR) modules, and the {\it Herschel}/PACS \oiiiup\ data (Fig.~\ref{fig:maps}) which, when combined with \oiiilw, can be used to derived the electron density (see Sect.~\ref{sec:ne}). 
While the \siii\ IRS maps have been taken from \cite{polles19} (see Sect. 3.1 of \cite{polles19} for the details of the data reduction), we re--reduced the PACS \oiiiup\, data using the transient correction pipeline available in HIPE\,15 \citep[see details in][]{fadda16} which calibrates the flux using the telescope background as absolute calibrator and corrects for transients. Comparing the reduced \oiiiup\, map with the \ha\, map (Sect.~\ref{sec:ha}) 
we noticed that the astrometry of PACS/\oiiiup\ was incorrect by a few arcseconds. Similar astrometric discrepancies have been seen in other PACS data, such as NGC~2146 and M100 \citep[see, e.g., ][]{fadda2023}. We corrected the PACS/\oiiiup\ astrometry by using the \ha\, map.

We complement these observations, which trace the gas properties, with dust continuum observations. The latter data are used to derive additional properties of the \hii\ regions, such as dust temperature. The dust emission of IC\,10 has been observed by {\it Spitzer}/MIPS and {\it Herschel}/PACS. In this paper we only consider the $24$\,$\mu$m, $70$\,$\mu$m and $100$\,$\mu$m bands. 
These data have been retrieved from the DustPedia (\citealt{davies17}) database\footnote{\url{http://dustpedia.astro.noa.gr}}. Since \ic\ is at low Galactic latitude, $b = -3.3^{\circ}$, the continuum maps are contaminated by galactic cirrus, requiring a careful foreground correction. For each photometric image we calculated the average foreground and the corresponding uncertainty ($\sigma_{\rm fg}$) taking the median and the median absolute deviation (MAD) of the lowest fluxes of the map, the `tail'. We identified the `tail' of the emission using an iterative procedure. First, we estimated the median and the MAD of all of the fluxes lower than an arbitrary limit, excluding the fluxes with a distance to the median higher than three MAD. Then we estimated the median again and the MAD of the remaining fluxes. This procedure has been performed until the value of the median converged (see Appendix~\ref{sec:background}). For each map, the corresponding contamination has been subtracted pixel by pixel. 

\subsubsection{\ha}\label{sec:ha}

The \ha\, map of \ic\, has been  obtained by \cite{hunterelme04} as part of a sample of 94 irregular galaxies. Several telescopes and instruments have been used to build the \ha\, catalogue of this sample. \ic\, has been observed using the Perkins Telescope at the Lowell Observatory at a spatial resolution of $2.2\arcsec$, and the reduced and continuum subtracted \ha\, map has been given to us by the authors. We refer to \cite{hunterelme04} for the details of the reduction. In this study we use the \ha\ map not corrected for extinction. 

\begin{deluxetable}{l c c c}
\label{tab:observations}
\tabcolsep=0.1cm
    \tablecaption{Characteristics of the observations.}
    \tablehead{\colhead{Tracer} &\colhead{Wavelength}&\colhead{FWHM}& \colhead{Pixel size}\\
    \colhead{} & \colhead{ ($\mu$m) } &\colhead{(arcsec)} & \colhead{(arcsec)}} 
        \startdata
        \noalign{\smallskip}
        H{\sc $\alpha$} & 0.6564 & 2.2 & 0.5 \\
	\noalign{\smallskip}
        [S\,\textsc{iii}] & 18.7 & 12 & 12 \\ 
	\noalign{\smallskip}
	[S\,\textsc{iii}] & 33.5 & 12 & 12 \\ 
        \noalign{\smallskip}
	[O\,\textsc{iii}] & 51.8 & 6.6 & 2\\
	\noalign{\smallskip} 
	[O\,\textsc{iii}] & 88.4 & 9.5 & 3\\
        \noalign{\smallskip}
        \hline
        \noalign{\smallskip}
        MIPS$\_$24 & 24 & 6 & 1.5 \\
        PACS$\_$70 & 70 & 5.6 & 2 \\
        PACS$\_$100 & 100 & 6.8 & 3 \\
	\noalign{\smallskip}
        \enddata
\end{deluxetable}

\begin{deluxetable}{l c c c c}
\label{tab:line_prop}
\tabcolsep=0.1cm
    \tablecaption{Properties of the infrared cooling lines used.}
    \tablehead{\colhead{Line} &\colhead{Wavelength}&\colhead{IP\,$^a$}&\colhead{$n_{\rm cr}^{b}$}&\colhead{$T_{\rm exc}$}\\
    \colhead{} & \colhead{ ($\mu$m) } &\colhead{(eV)}&\colhead{(cm$^{-3}$)}& \colhead{(K)}} 
        \startdata
        \noalign{\smallskip}
        [S\,\textsc{iii}] & 18.7 & 23.34 & 2$\times$10$^4$ & 769 \\
	\noalign{\smallskip}
	[S\,\textsc{iii}] & 33.5 & 23.34 & 7$\times$10$^3$ & 430 \\
        \noalign{\smallskip}
	[O\,\textsc{iii}] & 51.8 & 35.12 & 3.6$\times$10$^3$ & 440 \\
	\noalign{\smallskip} 
	[O\,\textsc{iii}] & 88.4 & 35.12 & 510 & 163 \\
	\noalign{\smallskip}
        \enddata
\tablecomments{ ($a$) Ionization potential, i.e., the energy necessary to create the ion; ($b$) Critical density for collision with electrons.}
\end{deluxetable}

\section{Electron density}\label{sec:ne}

We used two electron density diagnostics: the line ratio \oiiiup/\oiiilw\ tracing the electron density range between $10$ and $10^3$\,\cm, and the line ratio \siiiup/\siiilw\ tracing the electron density range between $100$ and $10^4$\,\cm. With the derived values, we investigated the distribution of the electron density inside the \hii\ regions of \ic.


\begin{figure*}
\centering
         \includegraphics[width=\textwidth, clip]{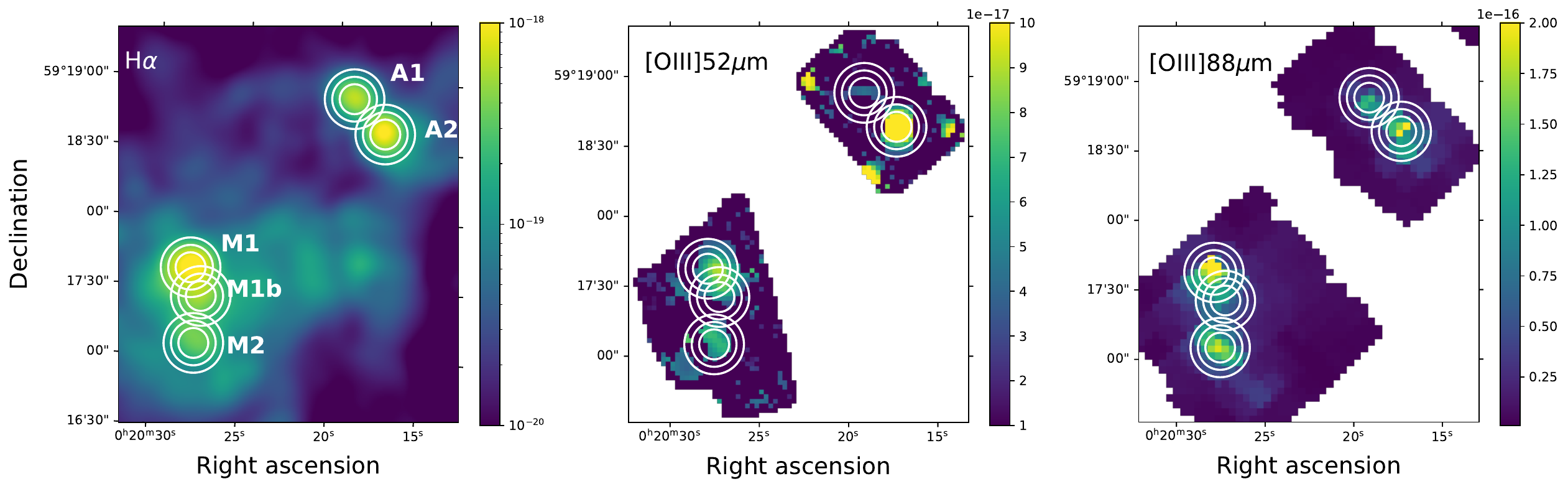}\\
         \includegraphics[width=.7\textwidth,clip]{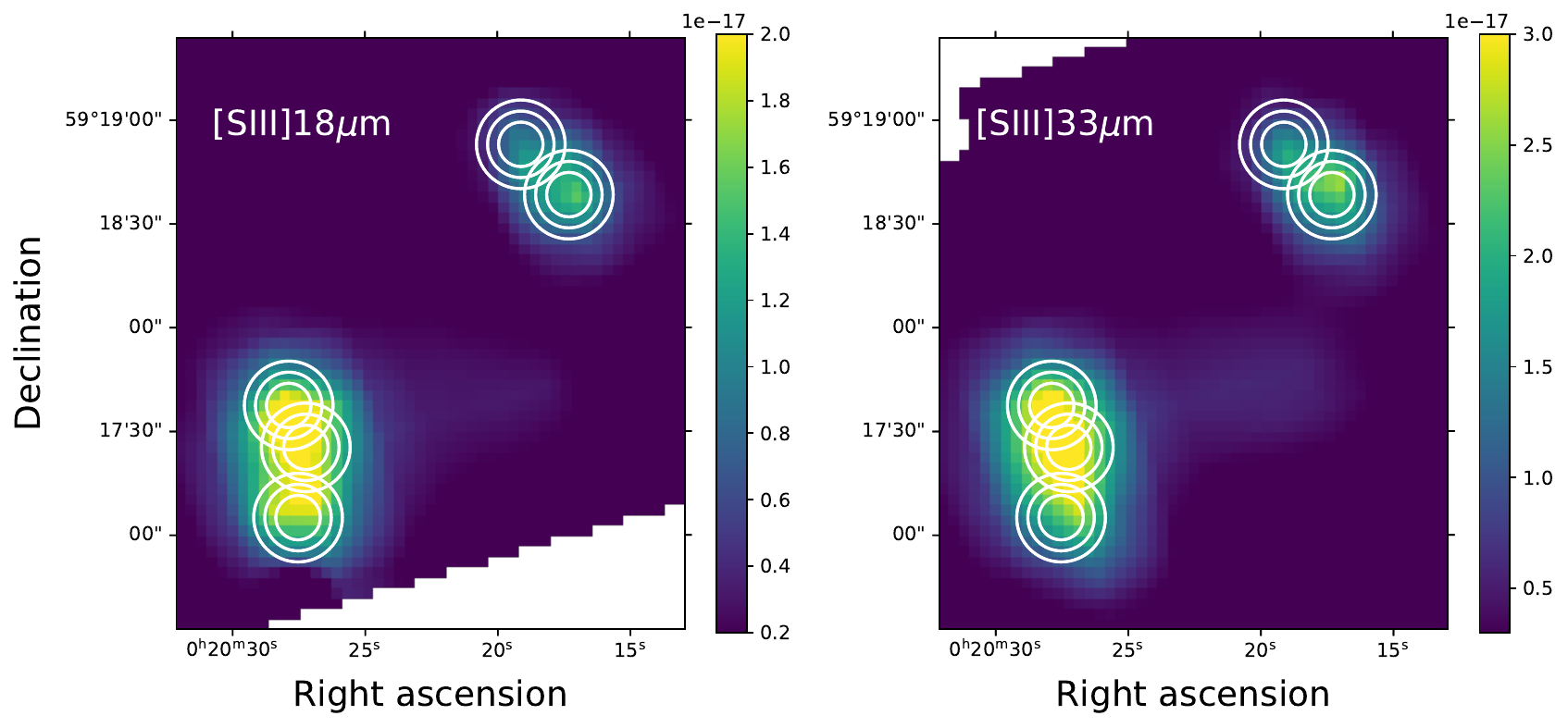}
    \caption{ The white circles indicating the regions and sub--regions described in Sect.\,\ref{sec:ne} overlap the intensity maps ($W\,m^{-2}pixel^{-1}$) of IC\,10. Top: \ha\, map (left), \oiiilw\, map (middle), \oiiiup\, map (right). All three maps are shown at the spatial resolution of $9.5\arcsec$. Bottom: \siiiup\, map (left) and \siiiup\, map (right), both shown at the spatial resolution of $12\arcsec$.}
    \label{fig:regions}
\end{figure*}

\subsection{Electron density using the diagnostic \oiii\ line ratios}\label{sec:neoiii}

The spatial resolution of the \oiii\, maps ($6.6\arcsec$ for \oiiilw\ and $9.5\arcsec$ for \oiiiup, see Table~\ref{tab:observations}) does not allow us to resolve individual \hii\, regions, therefore we identified larger regions, three of which are inside the main star-forming region of \ic, M1, M1b and M2, and two regions inside the arc, A1 and A2. For each region we calculated the electron densities, \neoiii, inside concentric circular areas in order to investigate the trends of the electron density. The selected circular regions with radii of $6.4\arcsec$ ($\sim$22.4 pc, c1), $9.6\arcsec$ ($\sim$36.6 pc, c2), and $12.8\arcsec$ ($\sim$44.8 pc, c3), are shown in Figure~\ref{fig:regions}. 

\begin{figure}[h!]
    \centering
        \includegraphics[width=\hsize]{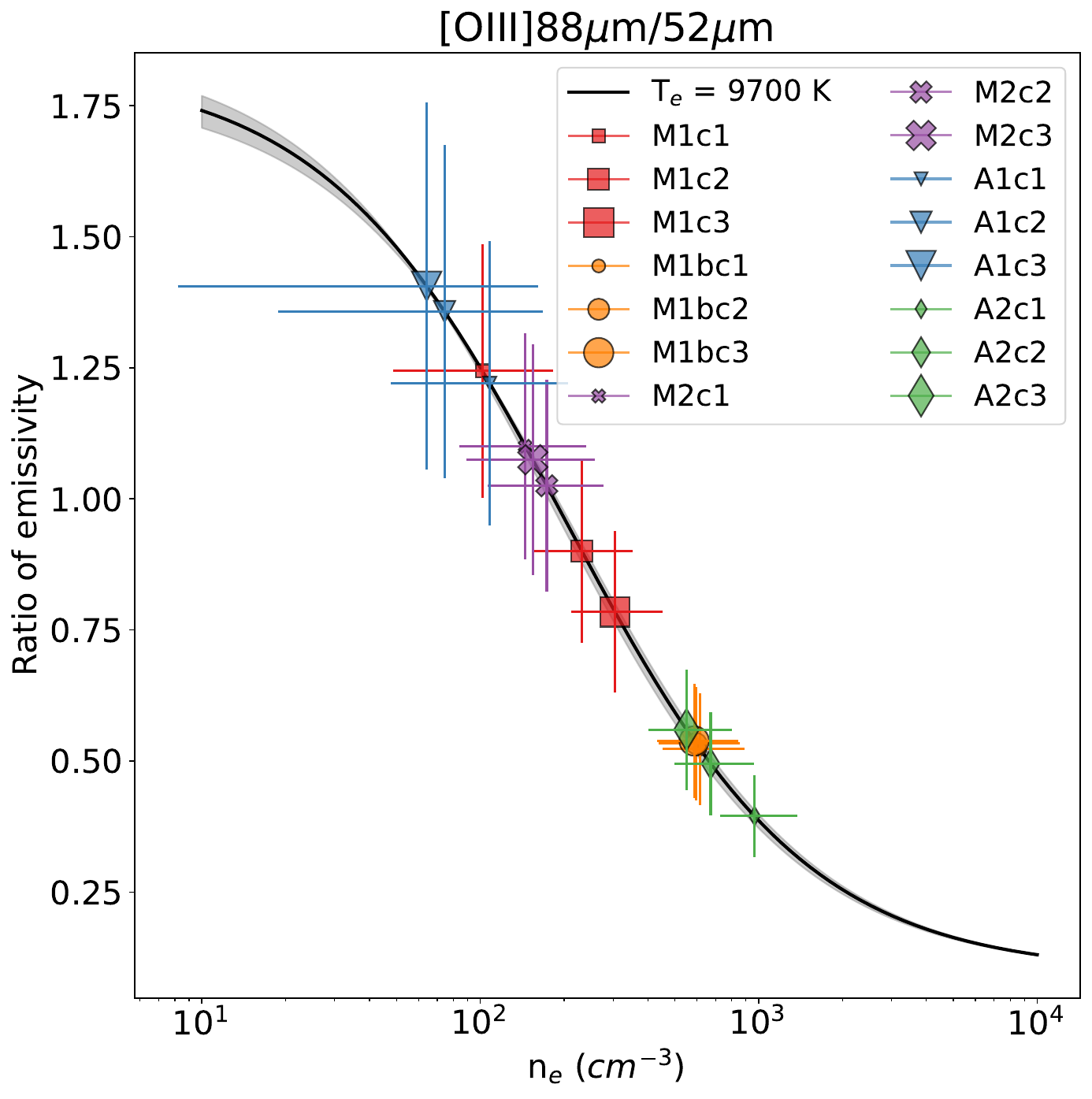}
    \caption{Theoretical ratios of \oiiiup/\oiiilw\ toward different regions of \ic\ (Figure~\ref{fig:regions}), determined at the temperature of 9700\,K as a function of the electron density n$_{\rm e}$. The different symbols indicate the observed line ratio values. Each color and symbol corresponds to one region, and the sizes of the symbol increase as the area considered for the integrated flux ratios increases. The gray are show the uncertainty due to the temperature's uncertainty: range between $10^{3.89}$~K to $10^{4.07}$~K.}
    \label{fig:ne_oiii}
\end{figure}

 The \oiiilw\ cube was convolved to the same spatial resolution of the \oiiiup\ cube. We then extracted the integrated spectrum in each circular aperture using SOSPEX \cite[]{fadda18, sospex}. We fitted the \oiiilw\ and \oiiiup\ emission lines with a Gaussian and a Voigt profile, respectively, assuming a constant value for the continuum. The fitting is weighted by the spectrum error. The integrated fluxes and the associated line fit uncertainties are presented in Table~\ref{tab:fluxesoiiisiii}. Once the fluxes were estimated, we calculated the \oiii\ ratio and derived the electron density, \neoiii, using the code PyNeb\footnote{\url{http://research.iac.es/proyecto/PyNeb}} (\citealt{luridiana15}) with its default atomic database\footnote{Several atomic database are available in PyNeb. The default atomic dataset is: PYNEB\_21\_01. The atomic data for O$^{++}$ are from \cite{FischerF2004} and Storey, P. J., Sochi, T.,  Badnell, N. R. 2014, MNRAS, 441, 3028; for S$^{++}$: Froese Fischer, C., G. Tachiev, and A. Irimia, 2006, At. Data Nucl. Data Tables 92, 607 cited by Podobedova, Kelleher, and Wiese 2009, J. Phys. Chem. Ref. Data, Vol.; Tayal  Gupta 1999, ApJ, 526, 544; for N$^+$ are from \cite{FischerF2004} ; Tayal 2011, ApJS, 195, 12.} (\citealt{morisset20}). While we assumed a temperature of 9700~K to calculate the electron density (see Appendix~\ref{sec:temperature}), the \oiii\ ratio is relatively insensitive to temperature, and therefore variations in temperature \cite[e.g.,][]{spinoglio15, kewley19b}. The variation in the predicted line ratios due to the uncertainty on the temperature is shown with the gray background in Figure~\ref{fig:ne_oiii}. The uncertainties of \neoiii\, are dominated by the uncertainties on the \oiii\ ratio, which have been estimated taking into account the line fit uncertainties and the calibration uncertainties ($15\%$ for \oiiilw\ and $12\%$ for \oiiiup). Table~\ref{tab:neoiiinesiii} presents the \neoiii\, values, while Figure~\ref{fig:ne_oiii} shows the derived \oiii\ ratio on the electron density curve.    

The derived \neoiii\ covers a range between 10~\cm\ to 1370~\cm. Even though the densities derived for each sub--region are compatible within the uncertainties, we can identify few trends of the \neoiii\ values as a function of the area. The region with the highest \neoiii\ values is A2. The \neoiii\ value of A2 decreases as the integrated area increases, indicating that at the spatial resolution of $9.5\arcsec$, A2 is a point source and increasing the integrated area we are including gas with lower density. The same behaviour is shown by the region A1, even though in this case high uncertainties are associated with the derived \neoiii\, values. The second densest region is M1b. The estimated \neoiii\ of this region and that of M2 are almost constant as the area is increased. The lack of a variation of \neoiii\, suggests that at least at the spatial resolution of $9.5\arcsec$, the regions M1b and M2 include \hii\, regions larger than the circle ``c1" and ``c2". Finally, the derived \neoiii\, values of M1 increase as the integrated area increases, showing an opposite trend to what we would expect in the presence of a compact \hii\, region. The behaviour shown by M1 suggests that increasing its area, we are including \hii\ regions with higher \neoiii\ than that of M1c1, rather than gas with lower density. 

\begin{deluxetable*}{l c c c c c c c}
\label{tab:fluxesoiiisiii}
\tabcolsep=0.1cm
    \tablecaption{Integrated fluxes.}
    \tablehead{\colhead{} &\colhead{} &\colhead{Region} &\colhead{} &\colhead{}&\colhead{Fluxes}& \colhead{} & \colhead{}\\
    \colhead{Name} &\colhead{R.A.} &\colhead{Dec.} &\colhead{radius} &\colhead{\oiiilw}&\colhead{\oiiiup} &\colhead{\siiilw}&\colhead{\siiiup}\\
    \colhead{} & \colhead{(J2000)} & \colhead{(J2000)} & \colhead{(arcsec)} & \colhead{ ($10^{-15}$\Wm) } &\colhead{($10^{-15}$\Wm)} & \colhead{ ($10^{-16}$ \Wm) } &\colhead{($10^{-16}$\Wm)}} 
        \startdata
        \noalign{\smallskip}
         M1c1 & 0:20:27.8835 & +59:17:37.523 & 6.4 &  1.97 $\pm$ 0.06 & 2.45 $\pm$ 0.02 & 3.46 $\pm$ 0.18 & 4.91 $\pm$ 0.36 \\
        \noalign{\smallskip}
         M1c2 & & & 9.6 & 4.16 $\pm$ 0.13 &  3.75 $\pm$ 0.02 & 6.59 $\pm$ 0.26 & 9.89 $\pm$ 0.54 \\
	\noalign{\smallskip}
	M1c3 & & & 12.8 & 6.65 $\pm$ 0.27 & 5.21 $\pm$ 0.02 & 10.16 $\pm$ 0.34 & 15.28 $\pm$ 0.68 \\
        \noalign{\smallskip}
         M1bc1 & 0:20:27.2390 & +59:17:25.400 & 6.4 & 2.16 $\pm$ 0.15  & 1.13 $\pm$ 0.01 & 3.21 $\pm$ 0.20 & 5.13 $\pm$ 0.49 \\ 
        \noalign{\smallskip}
         M1bc2 & & & 9.6 & 5.01 $\pm$ 0.33 & 2.67 $\pm$ 0.02 & 6.92 $\pm$ 0.30 & 10.60 $\pm$ 0.70 \\
	\noalign{\smallskip}
	M1bc3 & & & 12.8 & 8.79 $\pm$ 0.55 & 4.73 $\pm$ 0.02 & 11.51 $\pm$ 0.39 & 17.20 $\pm$ 0.87 \\
        \noalign{\smallskip}
         M2c1& 0:20:27.5330 & +59:17:05.082 & 6.4 & 1.62 $\pm$ 0.06 & 1.78 $\pm$ 0.01 & 2.54 $\pm$ 0.19 & 3.20 $\pm$ 0.39\\    
        \noalign{\smallskip}
         M2c2& & & 9.6 & 3.33 $\pm$ 0.15 & 3.41 $\pm$ 0.02  & 5.44 $\pm$ 0.28 & 7.32 $\pm$ 0.60 \\
	\noalign{\smallskip}
	M2c3 & & & 12.8 & 4.54 $\pm$ 0.31 & 4.88 $\pm$ 0.02 & 9.31 $\pm$ 0.37 & 12.61 $\pm$ 0.79 \\
        \noalign{\smallskip}
         A1c1& 0:20:19.1184 & +59:18:52.981 & 6.4 & 0.84 $\pm$ 0.09 & 1.02 $\pm$ 0.01 & 1.44 $\pm$ 0.17 & 2.61 $\pm$ 0.35\\
        \noalign{\smallskip}
        A1c2 & & & 9.6 & 1.24 $\pm$ 0.16 & 1.68 $\pm$ 0.01 & 2.96 $\pm$ 0.24 & 4.58 $\pm$ 0.47 \\
	\noalign{\smallskip}
	A1c3 & & & 12.8 & 1.79 $\pm$ 0.28 & 2.51 $\pm$ 0.02 & 4.67 $\pm$ 0.30 & 6.95 $\pm$ 0.58 \\
        \noalign{\smallskip}
        A2c1& 0:20:17.3062 & +59:18:38.423 & 6.4 & 4.71 $\pm$ 0.20 & 1.86 $\pm$ 0.01 & 2.09 $\pm$ 0.17 & 3.45 $\pm$ 0.37 \\
        \noalign{\smallskip}
         A2c2& & & 9.6 & 6.28 $\pm$ 0.32 & 3.11 $\pm$ 0.02 & 4.28 $\pm$ 0.24 & 7.08 $\pm$ 0.52 \\
	\noalign{\smallskip}
	A2c3 & & & 12.8 & 7.72 $\pm$ 0.56 & 4.32 $\pm$ 0.03 & 6.49 $\pm$ 0.31 & 10.72 $\pm$ 0.64 \\
        \noalign{\smallskip}
        \enddata
\end{deluxetable*}

\begin{deluxetable}{l c c}
\label{tab:neoiiinesiii}
\tabcolsep=0.5cm
    \tablecaption{Derived electron density for Te\,=\,9700\,K.}
    \tablehead{
    \colhead{Region} & \colhead{\neoiii}& \colhead{\nesiii}\\
    \colhead{} & \colhead{(\cm)} & \colhead{(\cm)}} 
        \startdata
        \noalign{\smallskip}
         M1c1 & 100$^{+80}_{-50}$ & 300$^{+100}_{-80}$\\
        \noalign{\smallskip}
         M1c2 & 230$^{+120}_{-80}$ & 250$^{+80}_{-70}$\\
	\noalign{\smallskip}
	M1c3 & 300$^{+150}_{-90}$ & 250$^{+70}_{-60}$ \\
        \noalign{\smallskip}
         M1bc1 & 610$^{+270}_{-160}$ & 200$^{+110}_{-80}$ \\ 
        \noalign{\smallskip}
         M1bc2 & 600$^{+260}_{-160}$ & 240$^{+90}_{-70}$ \\
	\noalign{\smallskip}
	M1bc3 & 590$^{+260}_{-150}$ & 260$^{+80}_{-60}$ \\
        \noalign{\smallskip}
         M2c1 & 140$^{+90}_{-60}$ & 400$^{+180}_{-130}$\\  
        \noalign{\smallskip}
         M2c2 & 170$^{+100}_{-70}$ & 340$^{+120}_{-90}$ \\
	\noalign{\smallskip}
	M2c3 & 150$^{+100}_{-60}$ & 340$^{+100}_{-80}$ \\
        \noalign{\smallskip}
         A1c1 & 110$^{+100}_{-60}$ & 120$^{+150}_{-100}$ \\
        \noalign{\smallskip}
         A1c2 & 70$^{+90}_{-60}$ & 230$^{+130}_{-90}$ \\
	\noalign{\smallskip}
	A1c3  & 60$^{+100}_{-50}$ & 260$^{+110}_{-90}$ \\
        \noalign{\smallskip}
        A2c1 & 960$^{+410}_{-240}$ & 180$^{+120}_{-90}$\\ 
        \noalign{\smallskip}
         A2c2 & 670$^{+290}_{-170}$ & 180$^{+90}_{-70}$ \\
	\noalign{\smallskip}
	A2c3 & 550$^{+250}_{-150}$ & 180$^{+80}_{-60}$ \\
        \noalign{\smallskip}
        \enddata
\end{deluxetable}

\subsection{Electron density using the diagnostic \siii\ line ratio}

\begin{figure}[h!]
    \centering
         \includegraphics[width=\hsize]{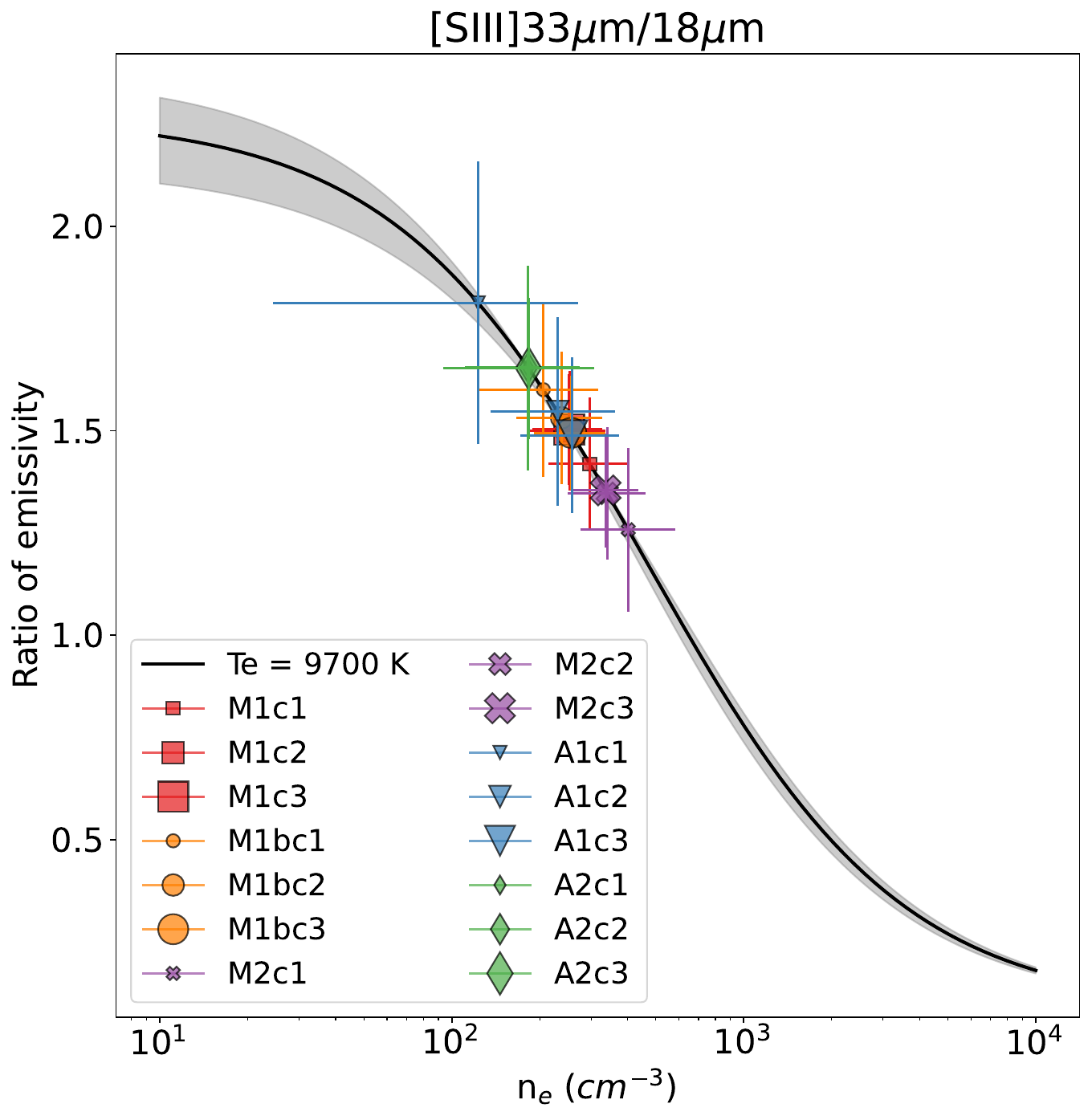}
    \caption{Theoretical ratio \siiiup/\siiilw\, at the temperatures of 9700\,K as a function of the electron density n$_{\rm e}$. The symbols indicate the observed line ratio values. The gray are show the uncertainty due to the temperature's uncertainty: range between $10^{3.89}$~K to $10^{4.07}$~K.}
    \label{fig:ne_siii}
\end{figure}

The integrated \siii\, fluxes have been calculated from the {\it Spitzer}/IRS LR maps. We integrated the fluxes using the same circular aperture used to calculate the \oiii\, integrated fluxes. For each integrated flux, the associated uncertainty is the square root of the sum of the squares of the uncertainty map values inside the region. The integrated fluxes and their uncertainties are presented in Table~\ref{tab:fluxesoiiisiii}. Once the fluxes were estimated, we calculated the \siii\ ratio and derived the electron density, \nesiii, using the code PyNeb and assuming a temperature of 9700~K. The uncertainties on the estimated \nesiii\ are due to the uncertainties on the ratio, while the uncertainties on the temperature are negligible. Figure~\ref{fig:ne_siii} shows the measured \siii\ ratio on the electron density curve, while Table~\ref{tab:neoiiinesiii} presents the derived \nesiii\ values.

The derived \nesiii\ values cover a range between 30~\cm\ to 570~\cm, narrower than that covered by \neoiii, and there is not a clear trend of \nesiii\ with the size of the aperture, for any of the regions. In light of the uncertainties on \nesiii, all of the derived \nesiii\ values are compatible with each other.

\subsection{Electron density stratification}

\begin{figure*}
    \centering
         \includegraphics[width=\textwidth]{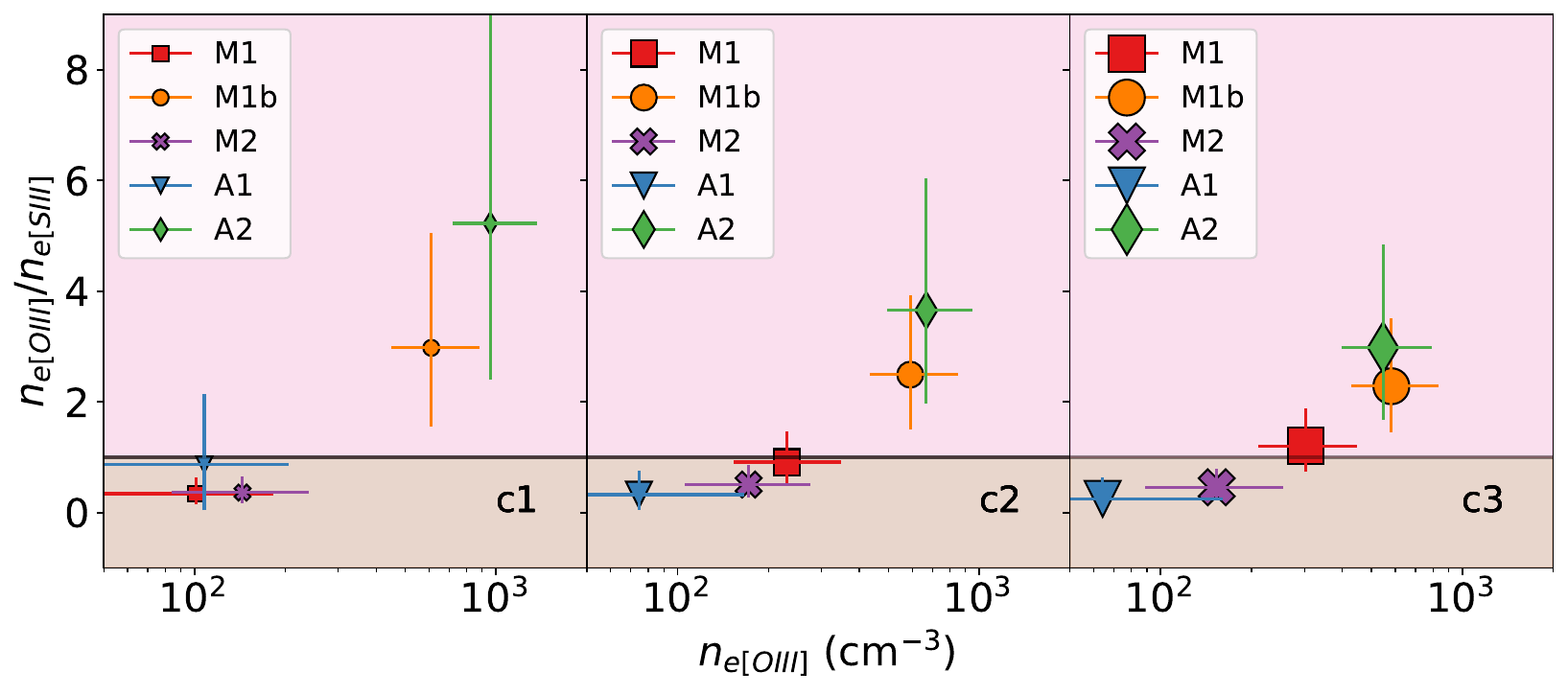}
    \caption{\neoiii/\nesiii\, ratio versus \neoiii\, values for different integrating areas (from left to right, c1, c2 and c3). The color code and symbols are the same as Figure~\ref{fig:ne_oiii}.}
    \label{fig:nesii_vs_neoiii}
\end{figure*}

Figure~\ref{fig:nesii_vs_neoiii} shows the comparison between \nesiii\ and \neoiii\ for each sub--region (c1, c2 and c3). The relation between \neoiii\ and \nesiii\ is not the same for all of the regions. We can identify two categories. (i) Regions with \neoiii\ always higher than \nesiii: M1b and A2, the two regions with the highest \neoiii. For these two regions,
the result suggests that the gas traced by the \siii\, ratio and the gas traced by the \oiii\, ratio arise from different components of the ionised gas. The highest density is traced by the ions with the highest ionization potential ($35.12$\,eV for the \oiii\ line vs $23.34$\,eV for \siii), suggesting that the densest clouds are closer to the ionizing source. (ii) Regions with \neoiii\,$\sim$\,\nesiii, within the uncertainties: M1, M2 and A1. In these cases, there is no clear density stratification, and both \oiii\ and \siii\ ratios trace gas of similar conditions.

\section{Discussion}\label{sec:discussion}

In the previous section we found that \nesiii\ is almost uniform between the \hii\ regions analyzed, while \neoiii\ covers a broad range. In the following we investigate if any of these behaviors correlates with the strength and the hardness of the radiation field of the \hii\ regions, and their evolutionary stages. The correlations are quantified using the Spearman's rank correlation coefficient calculated with the Python function $scipy.stats.spearmanr$.

\subsection{Electron density and radiation field}

\begin{figure*}[t!]
    \centering
         \includegraphics[width=\textwidth]{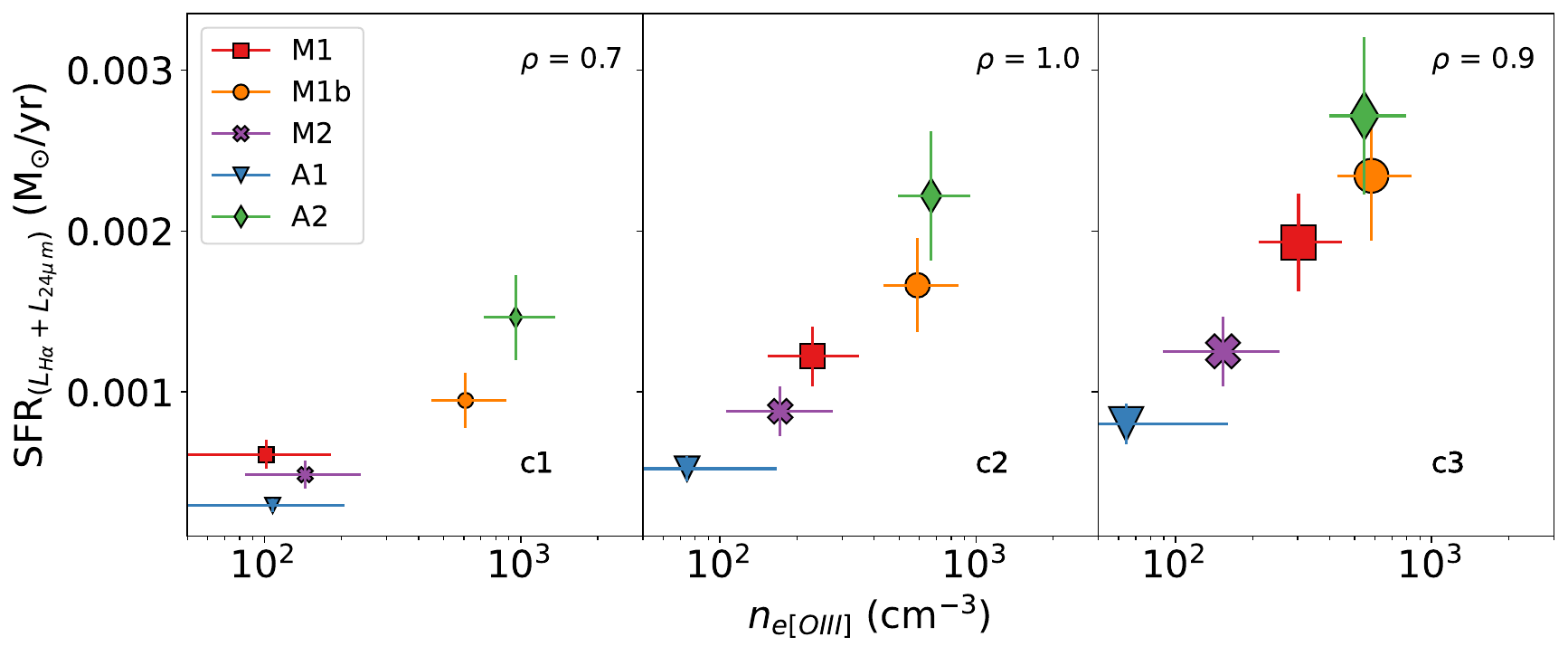}\\
         \includegraphics[width=\textwidth]{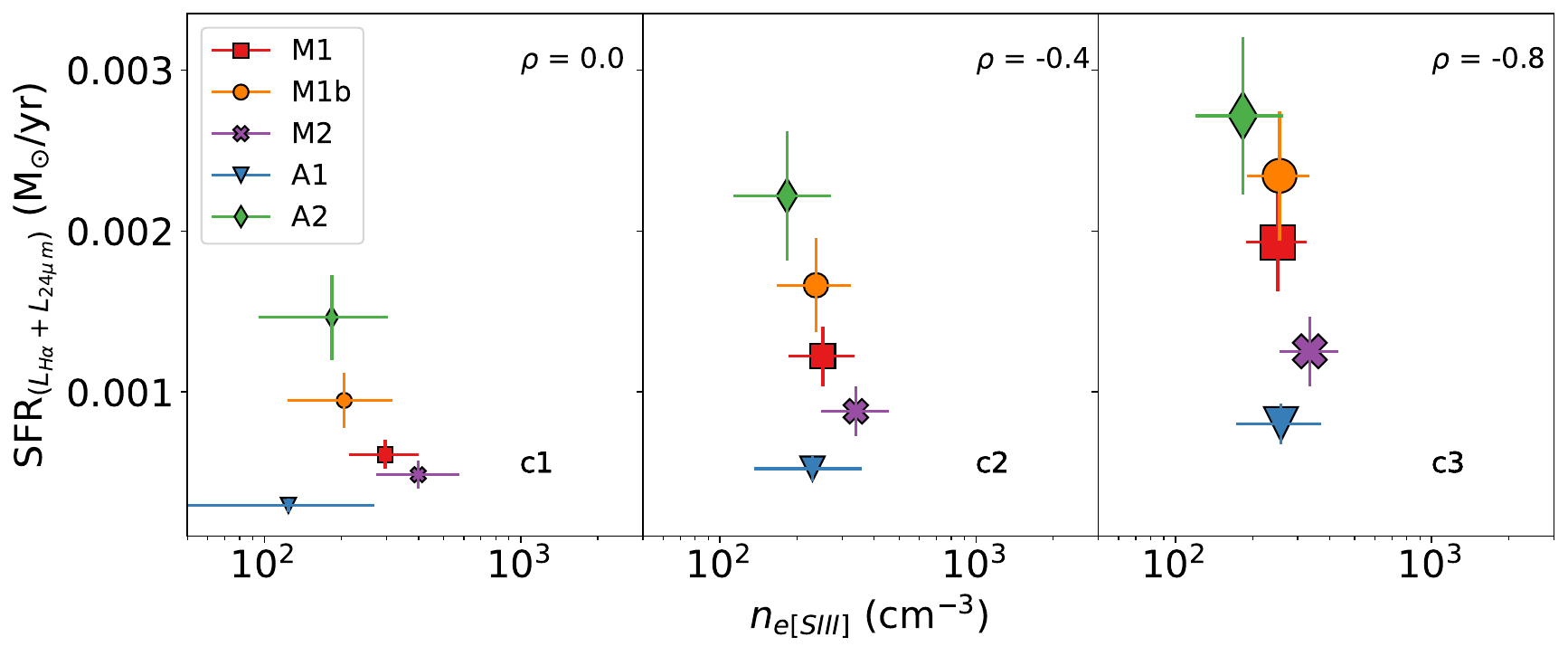}
    \caption{SFR versus \neoiii\ (top) and SFR versus \nesiii\ (bottom) for different integrating areas (from left to right, c1, c2 and c3). The color code and symbols are the same as Figure~\ref{fig:ne_oiii}. Spearman’s rank correlation coefficient are indicated in the top--right corner ($\rho$).}
    \label{fig:ne_SFR}
\end{figure*}

The combination of the optical line emission \ha\ and the dust emission at 24\,$\mu$m is one of the most reliable tracers of star formation rate (SFR; \citealt{kennicutt07}, \citealt{kennicutt09}). While both the \ha\ and  24\,$\mu$m are typically due to heating by young massive stars, they trace different evolutionary stages of the \hii\, regions. \ha\, is a good tracer of unobscured \hii\, regions, while the 24\,$\mu$m band is a good tracer of the ongoing dust-obscured SF. Thus, both \ha\ and 24\,$\mu$m are needed to estimate the total SF activity \citep{calzetti07}:
\begin{multline}\label{eqn:sfr}
    SFR (M_{\odot}yr^{-1}) = 5.3\times10^{-42}[L(H\alpha)_{obs}(ergs\,s^{-1}) \\+ (0.031\pm0.006)L(24\mu\,m)(ergs\,s^{-1})]
\end{multline}
where $L(H\alpha)_{obs}$ is the observed luminosity of H$\alpha$, not corrected for extinction. After convolving both maps to the spatial resolution of $9.5\arcsec$, we calculated the integrated fluxes of \ha\ and 24\,$\mu$m for each sub--region (see Table~\ref{tab:fluxes}) and calculated the corresponding SFR using equation~\ref{eqn:sfr}. Figure~\ref{fig:ne_SFR} shows a clear correlation between SFR and \neoiii, for all the sub--regions ($\rho$ equal to $0.7$, $1.0$, and $0.9$ for c1, c2 and c3, respectively), while the SFR does not seem to correlate ($\rho\,=\,0.0$) or it is anticorrelated ($-0.4$ and $-0.8$) with \nesiii. Given the uncertainties in the values of \nesiii, all of the values are consistent with one another. These results suggest that only the electron density traced by O$^{++}$, the ion with the highest ionization potential, is linked to the SFR: higher \neoiii\ could indicate higher SFR. 

\begin{figure*}[h!]
    \centering
         \includegraphics[width=\textwidth]{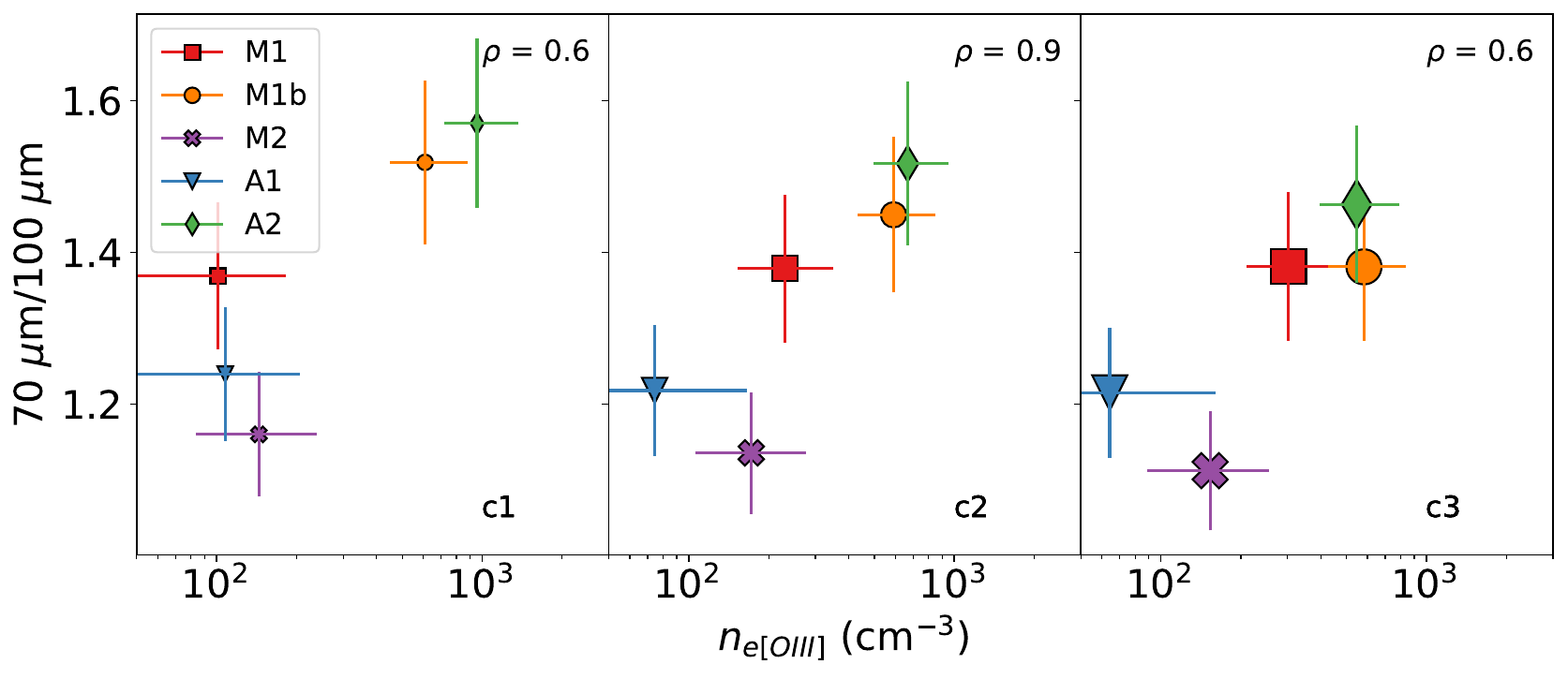}\\
         \includegraphics[width=\textwidth]{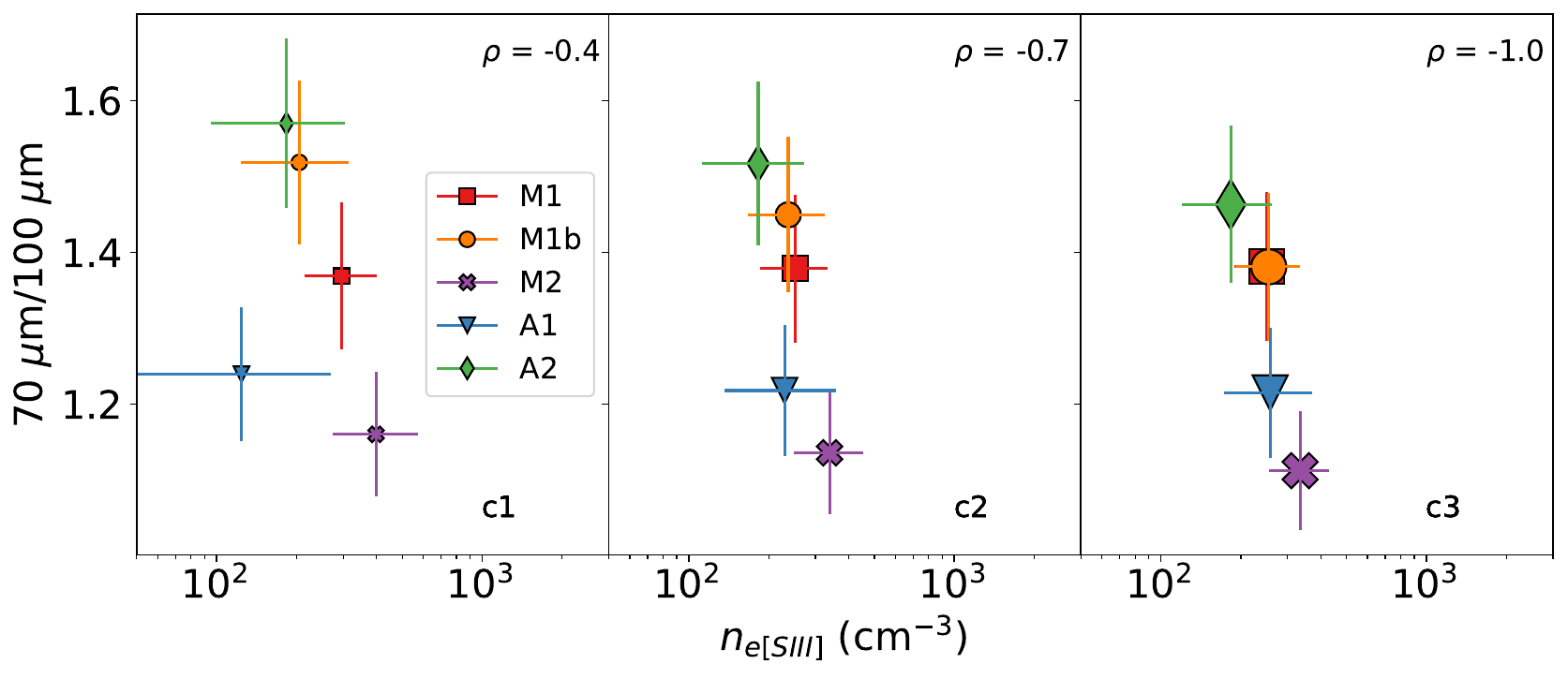}
    \caption{The dust continnum ratio $70$\,$\mu$m/$100$\,$\mu$m versus \neoiii (top), and versus \nesiii (bottom). The color code and symbols are the same as Figure~\ref{fig:ne_oiii}. Spearman’s rank correlation coefficient are indicated in the top--right corner ($\rho$).}
    \label{fig:ne_Te}
\end{figure*}

We can also investigate a possible correlation between the hardness of the radiation field,
the shape of the stellar spectrum, and \density, by examining an indicator of dust temperature, such as the dust continuum ratio $70$\,$\mu$m/$100$\,$\mu$m. The dust emission at shorter wavelengths becomes enhanced when the grains are exposed to harder interstellar radiation fields (ISRF, e.g., \citealt{madden06}). Therefore, we can use the ratio of two IR photometric bands such as $70$\,$\mu$m/$100$\,$\mu$m to trace the hardness of the radiation field and investigate its correlation with \neoiii\ and \nesiii. The integrated fluxes of the $70$\,$\mu$m and $100$\,$\mu$m continua have been estimated following the same procedure used to calculate the \siii\ fluxes, after convolving both maps to the spatial resolution of $9.5\arcsec$. Figure~\ref{fig:ne_Te} shows that $70$\,$\mu$m/$100$\,$\mu$m ratio correlates with \neoiii with a correlation coefficient of $0.6$, $0.9$, and $0.6$, for c1, c2 and c3, respectively. These correlations confirm that \neoiii\ is linked to the hardness of the radiation field. Again, given the uncertainties in the values of \nesiii, no trend in the $70$\,$\mu$m/$100$\,$\mu$m ratio with \nesiii can be reliably discerned.

Therefore, the electron density traced by the ions with high ionization potential, O$^{++}$, traces the properties of the radiation filed. The results of this study suggest that \neoiii\ reflects the density of the gas layer most affected by the stellar feedback processes and the evolutionary stage of the \hii\ regions, while the electron density of the second layer, traced by the \siii\ ratio, is not affected by the variation of the radiation field.

\begin{deluxetable*}{l c c c c}[!t]
\label{tab:fluxes}
\tabcolsep=0.1cm
    \tablecaption{Integrated fluxes.}
    \tablehead{\colhead{Region}&\colhead{\ha$^a$}&\colhead{24\,$\mu$m$^b$}&\colhead{70\,$\mu$m$^b$}&\colhead{100\,$\mu$m$^b$}\\
    \colhead{}&\colhead{(10$^{-16}$\Wm)}&\colhead{(10$^{-14}$\Wm)}&\colhead{(10$^{-13}$\Wm)}&\colhead{(10$^{-13}$\Wm)}} 
        \startdata
        \noalign{\smallskip}
         M1\,c1 & 5.30 & 4.38 $\pm$ 0.02 & 1.041 $\pm$ 0.004 & 0.761 $\pm$ 0.002\\
        \noalign{\smallskip}
         M1\,c2 & 9.34 & 9.14 $\pm$ 0.04 & 2.187 $\pm$ 0.007 & 1.587 $\pm$ 0.004 \\
	\noalign{\smallskip}
	M1\,c3 & 12.90 & 15.04 $\pm$ 0.05 & 3.682 $\pm$ 0.010 & 2.665 $\pm$ 0.005 \\
        \noalign{\smallskip}
         M1b\,c1& 2.83 & 8.52 $\pm$ 0.05 & 2.211 $\pm$ 0.011 & 1.456 $\pm$ 0.005 \\ 
        \noalign{\smallskip}
         M1b\,c2& 5.83 & 14.67 $\pm$ 0.06 & 3.817 $\pm$ 0.013 & 2.633 $\pm$ 0.007 \\
	\noalign{\smallskip}
	M1b\,c3 & 9.85 & 20.14 $\pm$ 0.06 & 5.344 $\pm$ 0.013 & 3.870 $\pm$ 0.007 \\
        \noalign{\smallskip}
         M2\,c1& 1.73 & 4.28 $\pm$ 0.01 & 1.542 $\pm$ 0.004 & 1.330 $\pm$ 0.003 \\  
        \noalign{\smallskip}
         M2\,c2& 3.33 & 7.68 $\pm$ 0.02 & 2.862 $\pm$ 0.005 & 2.520 $\pm$ 0.004 \\
	\noalign{\smallskip}
	M2\,c3 & 5.04 & 10.84 $\pm$ 0.02 & 4.192 $\pm$ 0.006 & 3.770 $\pm$ 0.005 \\
        \noalign{\smallskip}
         A1\,c1& 2.52 & 2.12 $\pm$ 0.01 & 0.670 $\pm$ 0.003 & 0.541 $\pm$ 0.002 \\
        \noalign{\smallskip}
         A1\,c2& 4.02 & 3.91 $\pm$ 0.01 & 1.201 $\pm$ 0.004 & 0.987 $\pm$ 0.003 \\
	\noalign{\smallskip}
	A1\,c3 & 5.15 & 6.30 $\pm$ 0.02 & 1.796 $\pm$ 0.006 & 1.478 $\pm$ 0.004 \\
        \noalign{\smallskip}
        A2\,c1& 4.02 & 13.26 $\pm$ 0.08 & 2.469 $\pm$  0.017 & 1.573 $\pm$ 0.009 \\ 
        \noalign{\smallskip}
         A2\,c2& 6.19 & 20.08 $\pm$ 0.09 & 3.830 $\pm$ 0.019 & 2.525 $\pm$ 0.011 \\
	\noalign{\smallskip}
	A2\,c3 & 7.72 & 24.53 $\pm$ 0.09 & 4.771 $\pm$ 0.019 & 3.261 $\pm$ 0.011 \\
        \noalign{\smallskip}
        \enddata
\tablecomments{($a$) Observed, no corrected. We assume a flux's uncertainty of 10$\%$. ($b$) The foreground uncertainties (4$\times$10$^{-19}$\Wm, 1$\times$10$^{-17}$\Wm\ and 2$\times$10$^{-17}$\Wm, for 24\,$\mu$m, 70\,$\mu$m and 100\,$\mu$m, respectively) are negligible in comparison with the uncertainties of the fluxes.}
\end{deluxetable*}

\subsection{Embedded \hii\ regions}

In this section we focus on the young \hii\ regions at early stages of their evolution: the compact \hii\, regions which are still embedded. We are not using \nesiii\ for this analysis, since from the previous section it is clear that this quantity does not correlate with other local physical properties of the \hii\ regions. 

\begin{figure*}[t!]
    \centering
         \includegraphics[width=\textwidth]{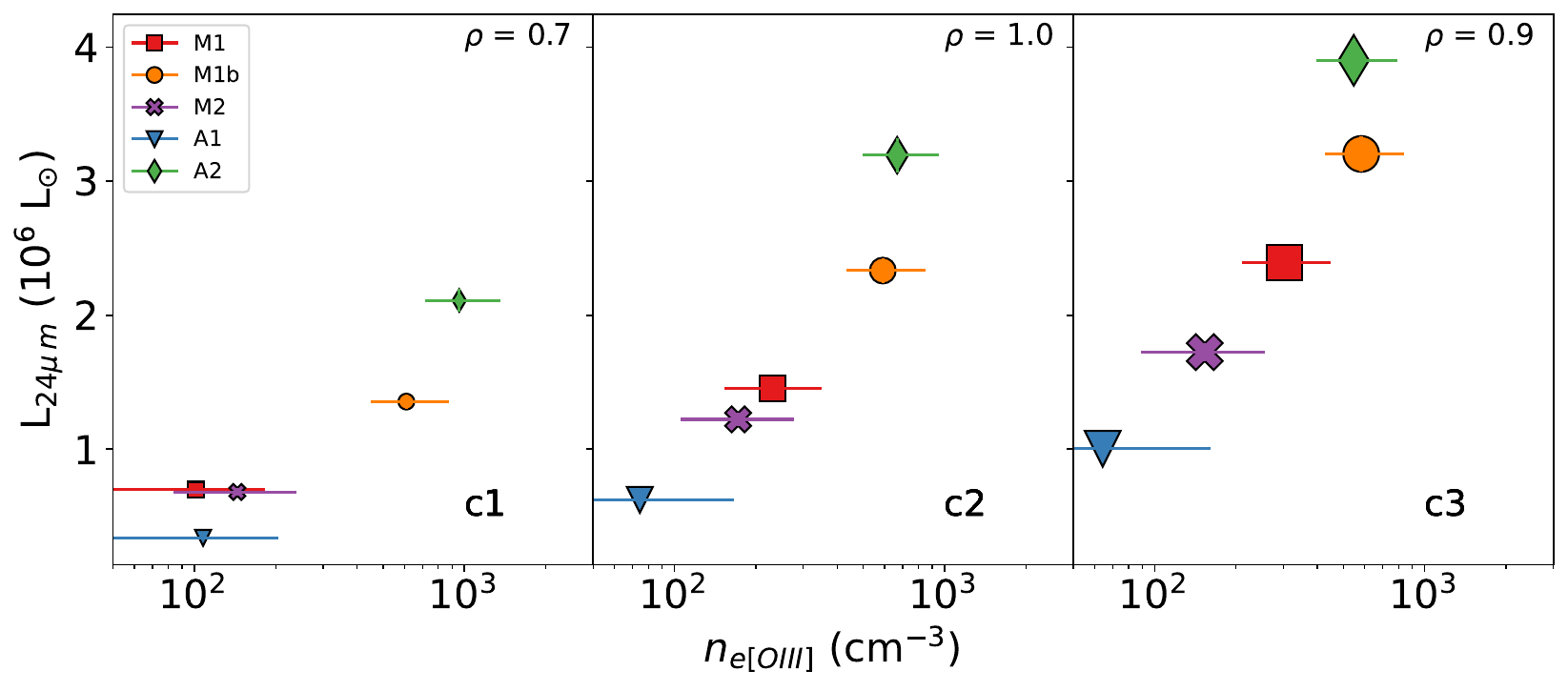}\\
         \includegraphics[width=\textwidth]{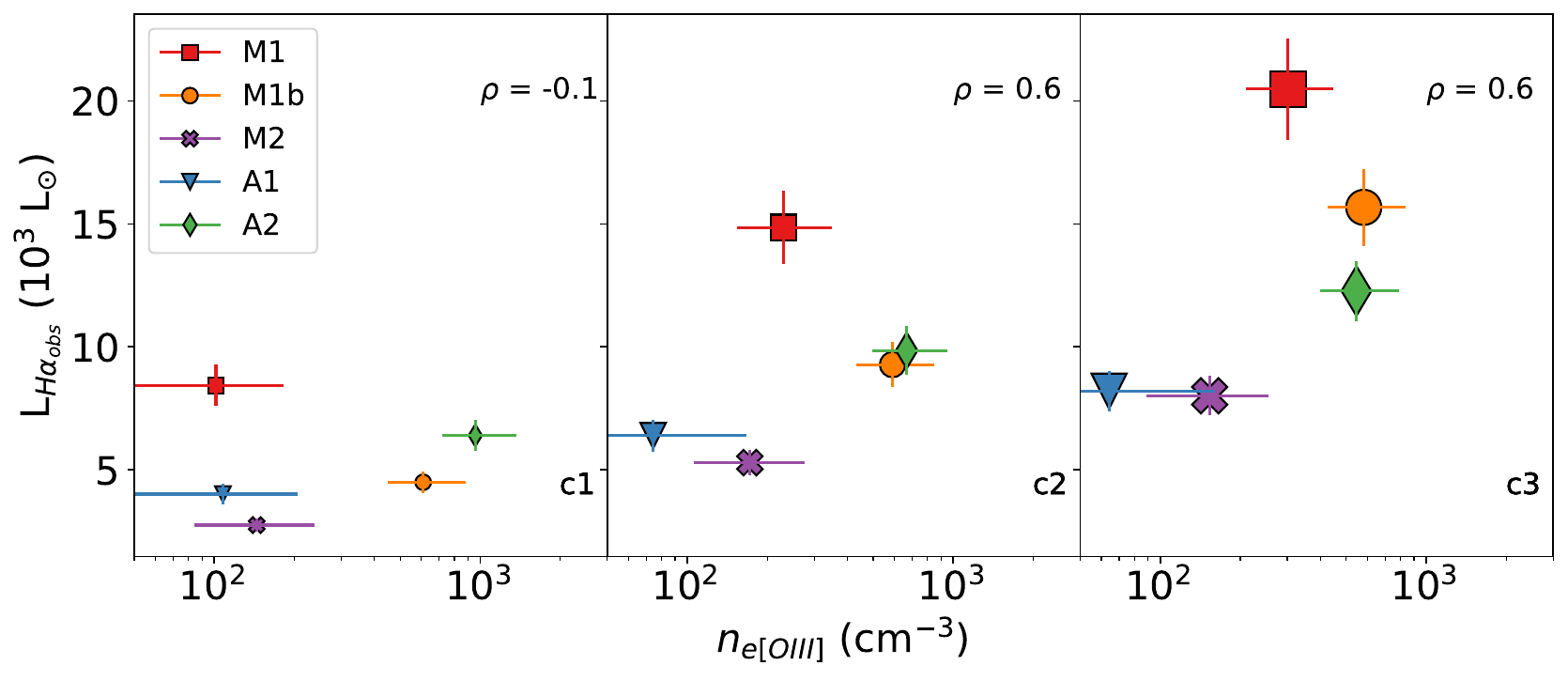}
    \caption{$24$\,$\mu$m luminosity versus \neoiii\, values (top) and observed H$\alpha$ luminosity versus \neoiii\, values (bottom). Left: c1 sub-regions; middle: c2 sub-regions; right: c3 sub-regions. Spearman’s rank correlation coefficient are indicated in the top--right corner ($\rho$).}
    \label{fig:24_vs_neoiii}
\end{figure*}

At the onset of the star formation, the ISM that has not yet been exposed to the stellar feedback is full of dust. These \hii\ regions are therefore bright in $24$\,$\mu$m but not in \ha, which is extincted by dust. In \cite{kim21}, for example, $24$\,$\mu$m has been used to measure the duration of the earliest embedded phase of star formation. Figure~\ref{fig:24_vs_neoiii} shows the $24$\,$\mu$m luminosity versus \neoiii\ (top panel), and the H$\alpha$ luminosity versus \neoiii\ (bottom panel). Each panel contains the results for one of the sub--regions (c1, c2 and c3). This figure highlights a clear correlation between the $24$\,$\mu$m luminosity and the \neoiii\, for all of the sub--regions ($\rho$ equal to 0.7, 1.0, and 0.9 for c1, c2, and c3, respectively), and a lower correlation between the observed H$\alpha$ luminosity and the \neoiii ($\rho\,=\,-0.1$ in the case of c1 and $\rho\,=0.6$ for c2 and c3). The brightest regions in 24\,$\mu$m are A2 and M1b, while the brightest region in the observed H$\alpha$ is M1. These results indicate that the densest \hii\, regions in \neoiii\ are also the dustiest. Moreover, the low correlation between H$\alpha$ and \neoiii\ suggests that the latter is a good tracer of the young embedded SF regions, while it is a less efficient tracer for evolved SF regions.

From the previous results, it is clear that the A2 region is a more extreme region than the other \hii\ regions of \ic. A2 stands out in that it is very bright in \oiiilw\, and in all of the dust emission tracers, especially at 24\,$\mu$m. In addition, the \neoiii\ value is much higher than those of the others \hii\, regions. These characteristics suggest that A2 is an embedded dense \hii\, region. This conclusion is confirmed by the radio continuum emission. A2 is the brightest source with e-MERLIN observations in 5 and 1.5 GHz by \cite{westcott17}. The authors classified A2 as a compact \hii\, region, combining 5 and 1.5 GHz data with \ha\, and 70\,$\mu$m (see Figure\,3 of \citealt{westcott17}). 

\begin{figure}[h!]
    \centering
         \includegraphics[width=\hsize]{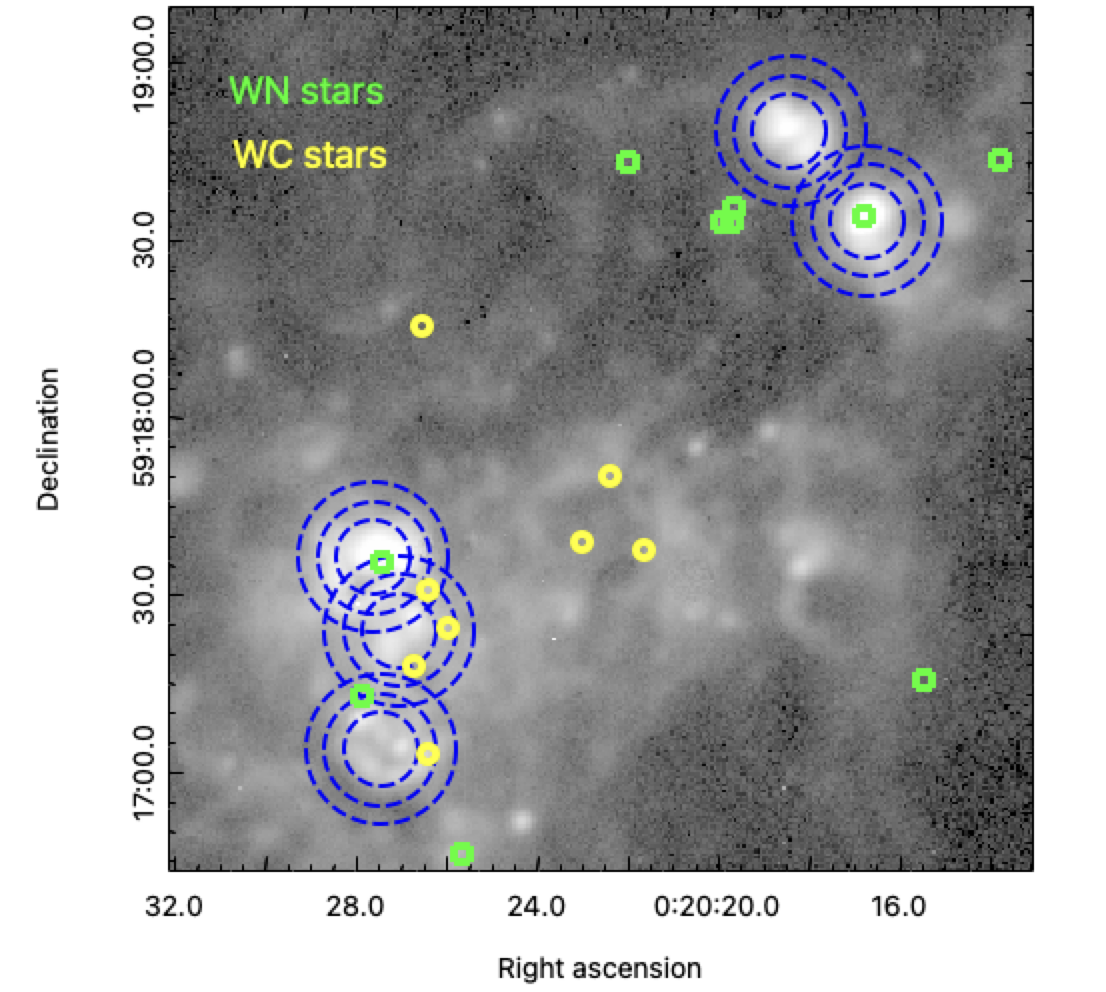}
    \caption{\ha\, image of IC\,10 with the positions of the Walf-Rayet stars from \cite{tehrani17} overlaid. The regions analyzed in this study are indicated with blue circles.}
    \label{fig:wr}
\end{figure}

The young age of the \hii\, region A2 is also supported by the presence of an early nitrogen sequence (WNE) type of WR star at its center (see Figure~\ref{fig:wr}). The WR stars are the post-main-sequence stage of the high mass O stars, and the WNE is the first phase of a WR. All of this leads to a picture of A2 as a young embedded dense \hii\, region with a massive WNE at its center, which is too young to have dispersed the surrounding dust with its feedback. Higher angular resolution data (e.g. GMOS data, \citealt{tehrani17}; KCWI data, \citealt{cosens22}) show that, among the several WR stars of \ic\ (see Figure~\ref{fig:wr}), three of these are situated inside an \hii\ region: the aforementioned WNE at the center of A2, the O2.5~If/WN6 inside M1c1, and the WC4 inside M2c2. The KCWI data shows that the O2.5~If/WN6 star is located at the edge of the \hii\ region (Figure 26 of \cite{cosens22}). The notation O2.5~If/WN6 indicates a star in the intermediate stage between an O2.5 and a WN6 star. The young evolutionary stage of this WR star and its position suggest that this star is a product of a second generation of star formation associated with the evolved \hii\ region complex inside M1c1. This scenario would agree with the bright emission in \ha\ and the low emission in 24\,$\mu$m of M1c1. The WC4 star inside M2c2 is located at the center of a compact \hii\ region that is visible with high spatial resolution (Figure 26 of \citealp{cosens22}), but with emission diluted at the spatial resolution of c2 ($19.6\arcsec$). Moreover, the WC4 designation indicates a WR star that is more evolved than a WNE star. Therefore, A2 is unique in hosting an early stage WN at its center.

\section{Summary}\label{sec:conclusion}
We combined our \oiiilw\, SOFIA/FIFI-LS data with {\it Herschel}/PACS \oiiiup\, data, and the \siiilw\, and \siiiup\, data from {\it Spitzer}/IRS to investigate the electron density of the brightest \hii\, regions of the nearby dwarf galaxy \ic\ at different spacial resolution: 44.8~pc (c1), 73.2~pc (c2), and 89.6~pc (c3).

The range of values covered by the derived \neoiii\ is broad, between 10~\cm\ and 1370~\cm. For the regions A1 and A2, the derived \neoiii\, decreases when considering larger areas, suggesting that those are point sources, at the spatial resolution of $9.5\arcsec$. In the case of M1, \neoiii\ increases when the area is enlarged. This different behaviour can be explained by the fact that the M1 region is made up of several \hii\, regions. We identified the center of the \hii\, regions based on the \ha\, luminosity, and in the case of M1, it seems that the position of the peak in \ha\ emission does not correspond to the densest \hii\ region of the area. 

The electron densities derived using the \siii\ ratio, are more uniform than those derived from \oiii\ ratio. The \nesiii\ values are between 30~\cm\ and 570~\cm. Comparing \neoiii\ with \nesiii, we found that for two regions, M1b and A2, the \oiii\ and the \siii\ ratios trace different components of the ionised gas. While for the other regions, i.e. M1, M2, and A1, \neoiii $\sim$ \nesiii\ across the different sub--regions. These results indicate that the distribution of the gas components inside the \hii\ region is not straightforward. Our study suggests that the electron density distribution gives us information about the evolutionary stage of the \hii\, region. In the case of A2, a confirmed embedded compact \hii\, region hosting a WNE star at the center, the highest density is traced by the \oiii\ line ratio, with O$^{++}$ being the ion with the highest ionization potential (35 eV), indicating that the densest clouds are closer to the ionizing source. On the other side, \neoiii\ $\sim$ \nesiii\, in unobscured \hii\, regions such as M1. This hypothesis is supported by the discovery that \neoiii\ correlates with the 24\,$\mu$m emission, which traces the embedded young SF activity.

Finally, we found that \neoiii\, correlates with SFR and dust temperature. Thus, the electron density traced by \oiii\, is also an indicator of the hardness and the intensity of the radiation field, with higher \neoiii\, found in more energetic environments, while the gas layers traced by \siii\, do not depend on the properties of the radiation field.

\ic\ has been the pilot galaxy to study the electron density stratification inside \hii\ regions using the infrared \siii\ and \oiii\ line ratios. To verify if the conclusions inferred studying the \hii\ regions of IC\,10 are valid in general, it will be necessary to expand the analysis to a heterogeneous sample of nearby galaxies, thereby providing a quantitative understanding of the environmentally--dependent interplay between stars and gas. Galaxies that have been observed with integral field unit 
(IFU) such us VLT/MUSE (\citealt{bacon10}), CFHT/SITELLE (\citealt{drissen19}), or JWST/MIRI (\citealt{rieke15}) would be the ideal objects for these followup studies.

\section{Acknowledgments}
We wish to thank the anonymous referee for providing insightful comments, which helped to improve this paper.
This paper was based on observations made
with the NASA/DLR Stratospheric Observatory for Infrared
Astronomy (SOFIA) and with Herschel. SOFIA was jointly
operated by the Universities Space Research Association, Inc.
(USRA), under NASA contract NNA17BF53C, and the
Deutsches SOFIA Institut (DSI) under DLR contract 50 OK
0901 to the University of Stuttgart.
MC and LR gratefully acknowledge funding from the DFG through an Emmy Noether Research Group (grant number CH2137/1-1).
COOL Research DAO is a Decentralized Autonomous Organization supporting research in astrophysics aimed at uncovering our cosmic origins.

%

\vspace{5mm}
\facilities{SOFIA (FIFI-LS), Herschel (MIPS, PACS), Spitzer (IRS),Perkins Telescope at the Lowell Observatory}


\software{astropy \citep{astropy13,astropy18,astropy22},  
          Cloudy \citep{ferland13}, 
          SOSPEX \citep{fadda18,sospex}
          }



%
\bibliography{biblio}{}
\bibliographystyle{aasjournal}
%

\appendix

\section{ Estimate of the foreground cirrus contamination}\label{sec:background}
\begin{figure*}[b!]
    \centering
         \includegraphics[width=0.5\textwidth]{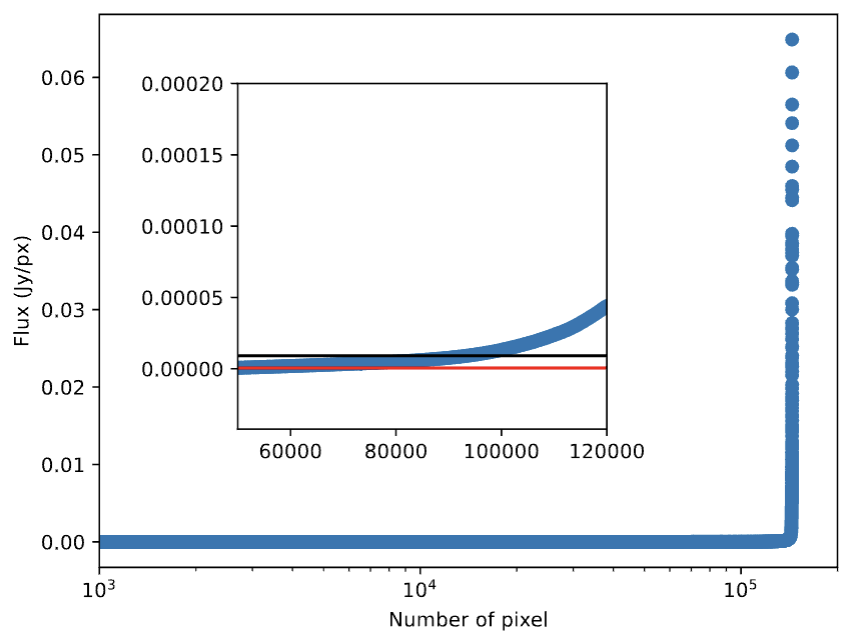}\includegraphics[width=0.5\textwidth]{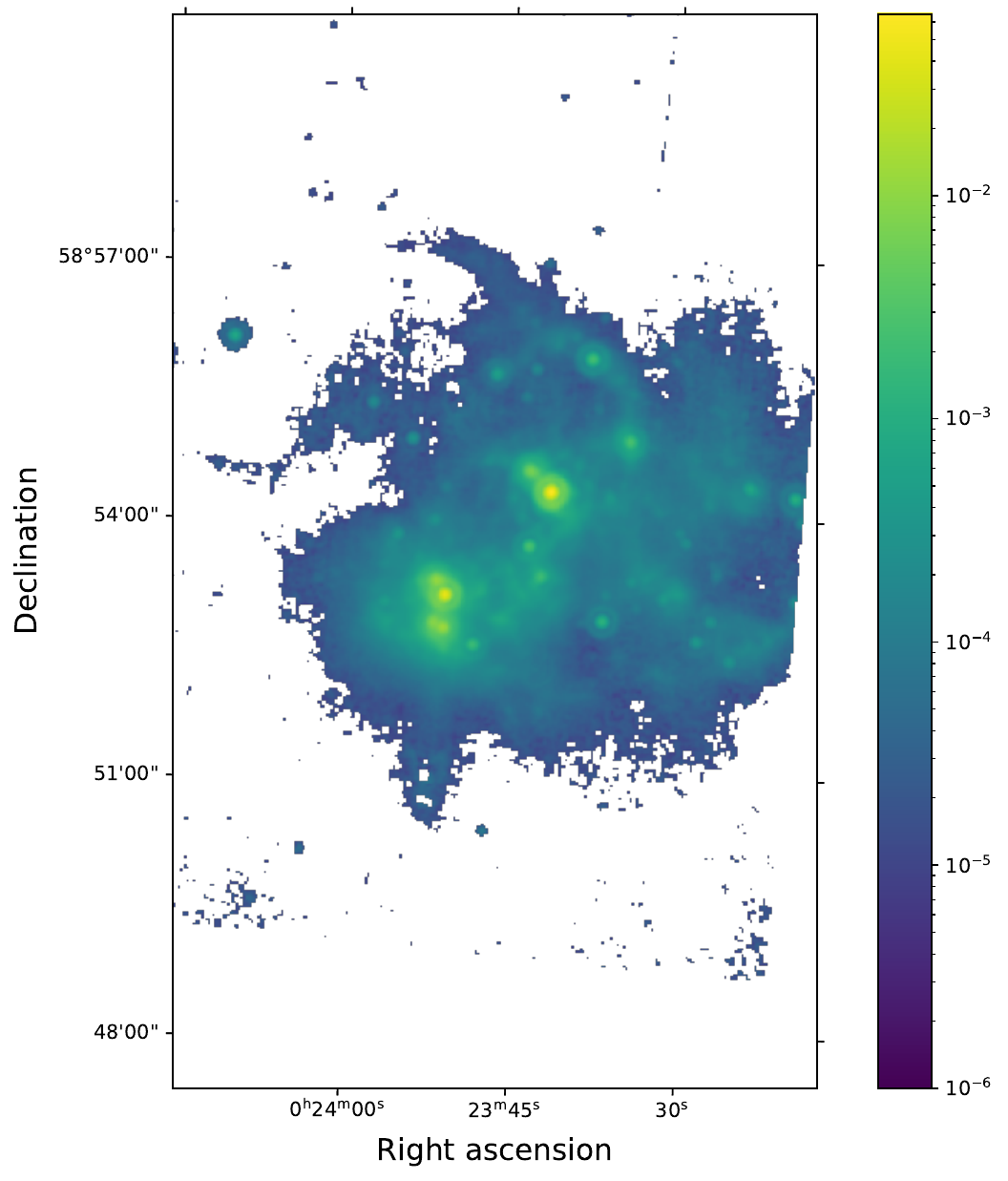}
    \caption{Left: Increasing distribution of the 24\,$\mu$m fluxes, with a zoom on the distribution showing the point where the emission starts to increase rapidly. The black line shows the upper limit of the fluxes used to calculate the foreground emission, while the red line corresponds to the derived foreground emission. Right: map of the continuum 24\,$\mu$m emission showing only the pixels with a flux higher than the upper limit used to identify the `tail'(black line on the plot on the left).}
    \label{fig:fg_24}
\end{figure*}

\begin{figure*}[t!]
    \centering
         \includegraphics[width=0.5\textwidth]{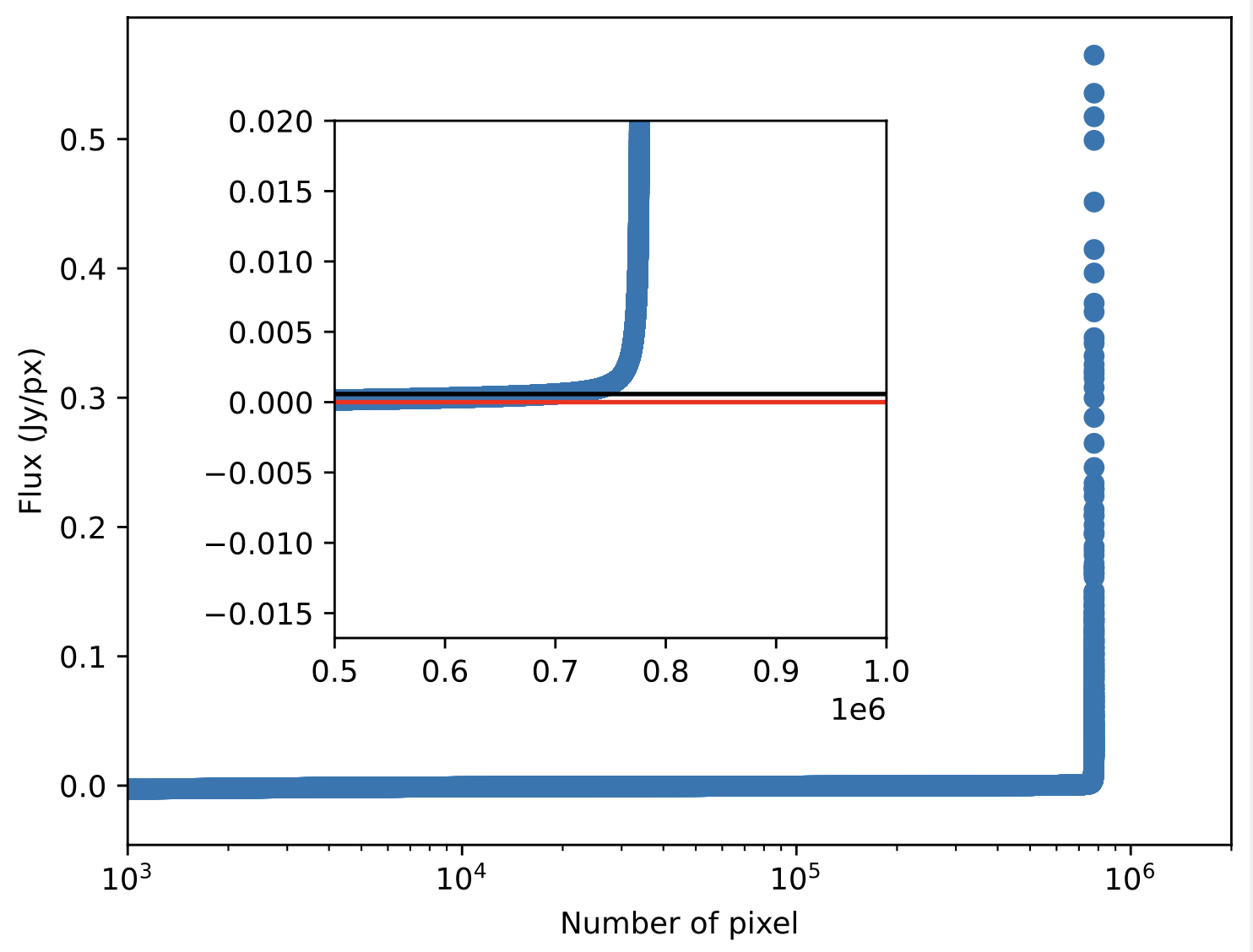}\includegraphics[width=0.5\textwidth]{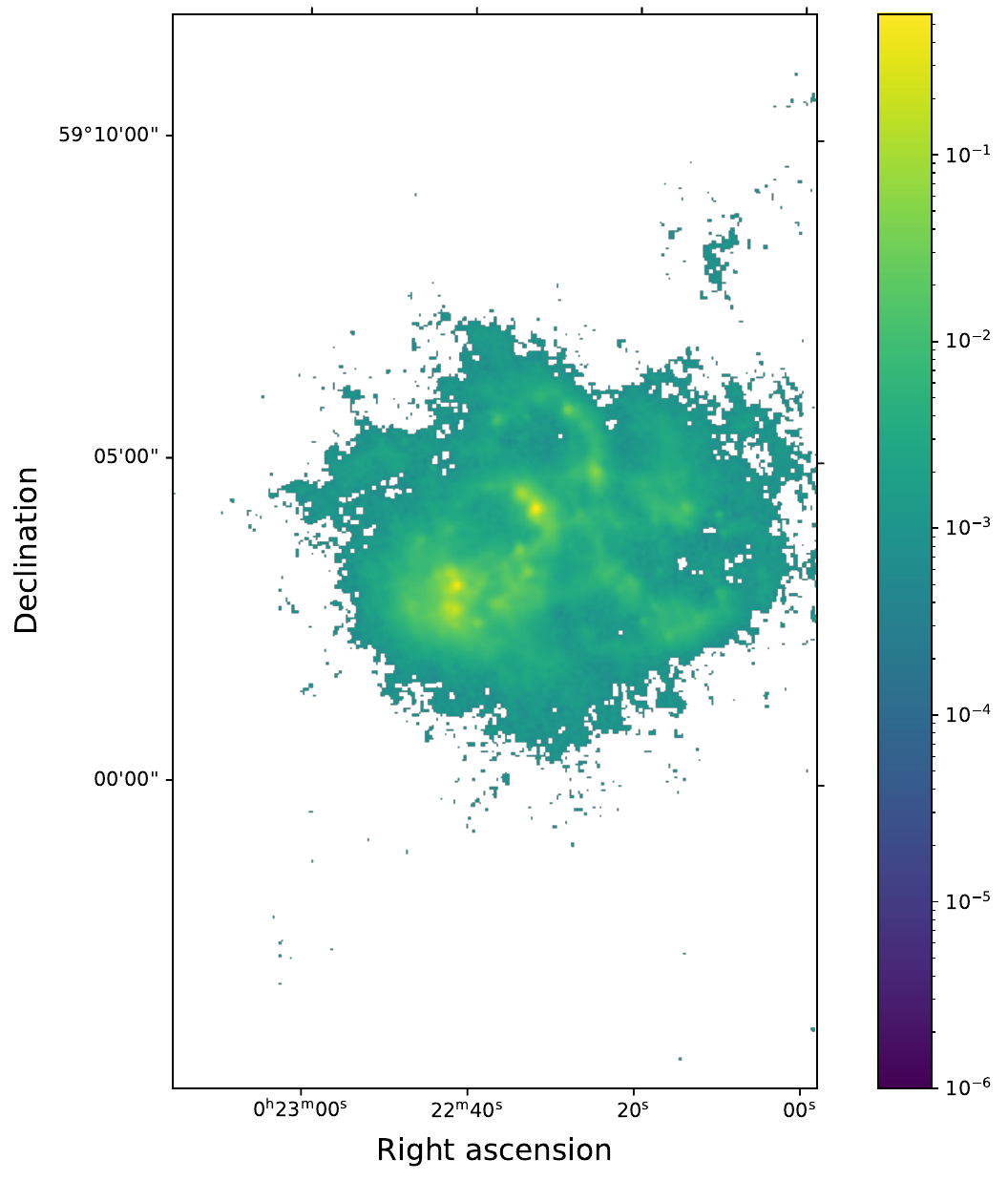}
    \caption{Left: Increasing distribution of the 70\,$\mu$m fluxes, with a zoom on the distribution showing the point where the emission starts to increase rapidly. The black line shows the upper limit of the fluxes used to calculate the foreground emission, while the red line corresponds to the derived foreground emission. Right: map of the continuum 70\,$\mu$m emission showing only the pixels with a flux higher than the upper limit used to identify the `tail' (black line on the plot on the left).}
    \label{fig:fg_70}
\end{figure*}

\begin{figure*}
    \centering
         \includegraphics[width=0.5\textwidth]{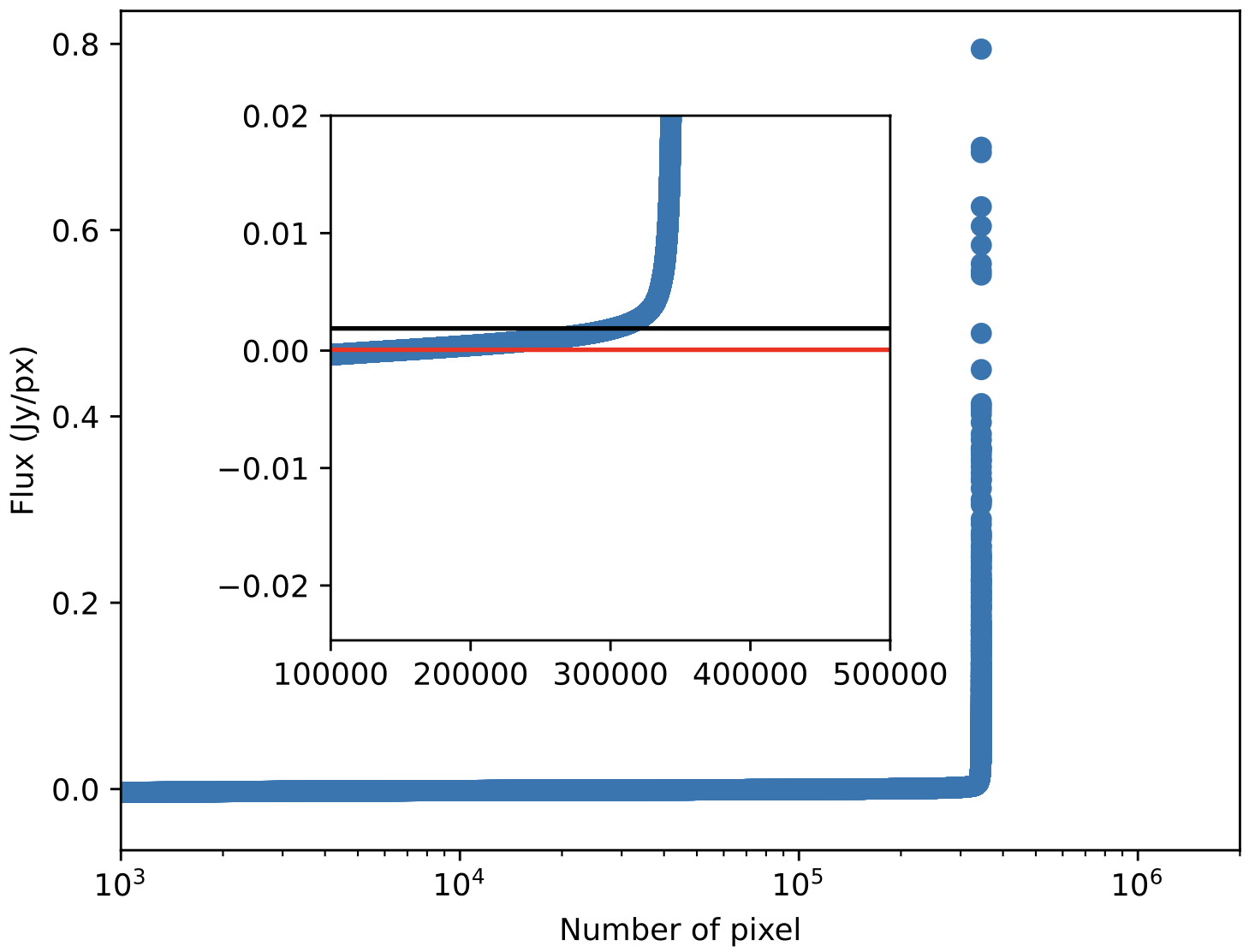}\includegraphics[width=0.5\textwidth]{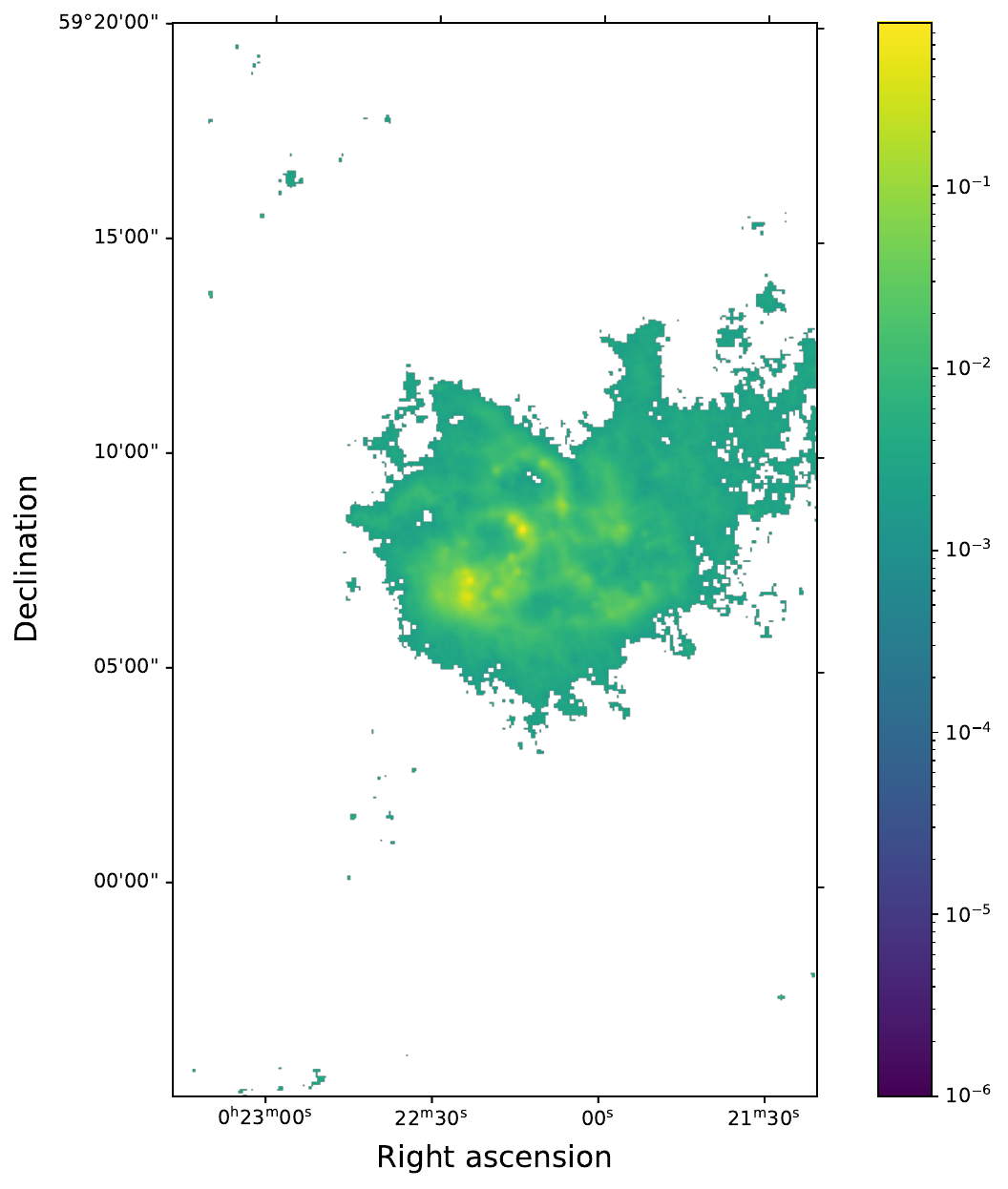}
    \caption{Left: Increasing distribution of the 100\,$\mu$m fluxes, with a zoom on the distribution showing the point where the emission starts to increase rapidly. The black line shows the upper limit of the fluxes used to calculate the foreground emission, while the red line corresponds to the derived foreground emission. Right: map of the continuum 100\,$\mu$m emission showing only the pixels with a flux higher than the upper limit used to identify the `tail' (black line on the plot on the left).}
    \label{fig:fg_100}
\end{figure*}

Figure~\ref{fig:fg_24}, left panel, shows the increasing distribution of all of the fluxes of the continuum 24\,$\mu$m map. The small window inside the plot shows the zoom on the section of the distribution where the emission starts to increase drastically. The upper limit of the final `tail' identified with the iterative procedure described in Sec.~\ref{sec:data} is shown with a black line, while the red line correspond to the median value, i.e. foreground emission. The right panel of Figure~\ref{fig:fg_24} shows the pixels with a flux value higher than the upper limit of the `tail'. Fig.~\ref{fig:fg_70} and Fig.~\ref{fig:fg_100} show the same as Figure~\ref{fig:fg_24} but for the continuum 70\,$\mu$m and 100\,$\mu$m, respectively.

\section{Temperature variation}\label{sec:temperature}
\begin{figure}[t!]
    \centering
         \includegraphics[width=0.5\textwidth]{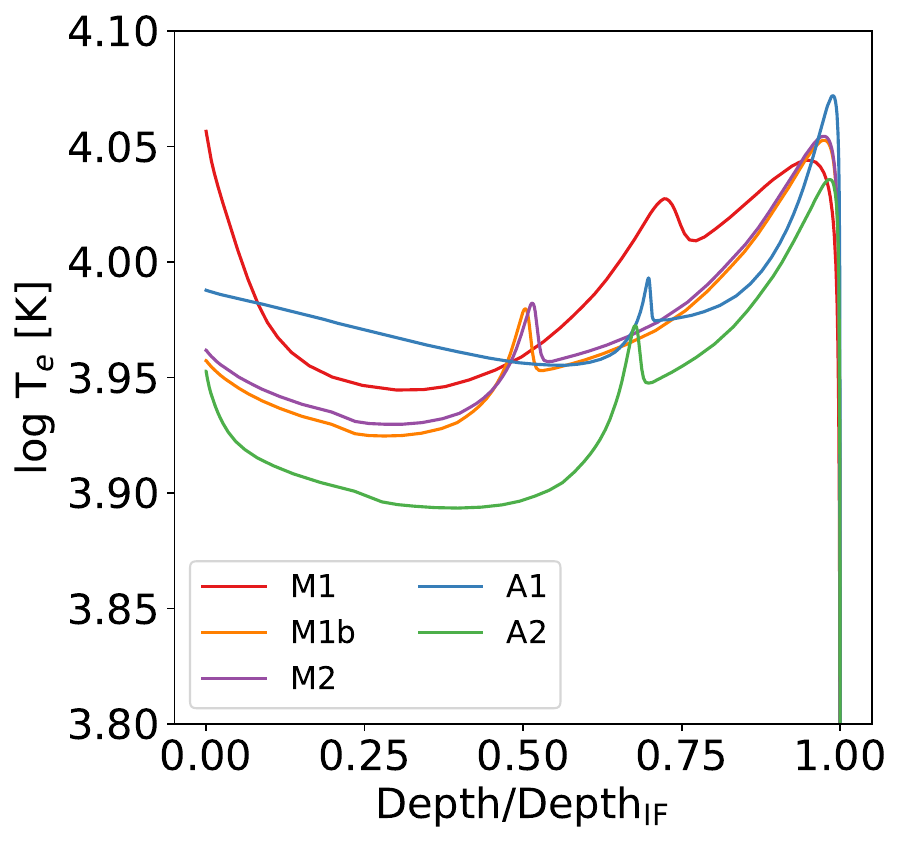}
    \caption{Gas temperature profiles calculated by the best-model solution of each of the \hii\, region studied. The depth is normalized to the value reached at the ionization front (Depth$_{\rm IF}$).}
    \label{fig:temperature}
\end{figure}

The low energy of the transition of the IR fine-structure levels compared to the gas temperature makes these ratios relatively insensitive to temperature, except in the low--density limit \cite[e.g.,][]{osterbrock06, kewley19}. The IR \oiii\, ratio and the \siii\, ratio are sensitive to the temperature variation for \density$<10^{1.9}$ and \density$<10^{2.5}$, respectively. Since the gas temperature is affected by the metallicity, with colder temperature for an \hii\, region with higher metallicity, the same dependency of the derived \neoiii\, and \nesiii\, seen for the gas temperature can be seen for the metallicity (see Figures~3, 15 and 16 of \citealt{kewley19b}). Deriving the exact temperature corresponding to the metallicity requires a complete photoionization modeling. Since those models have been already performed and presented in \cite{polles19} for each investigated region (i.e. M1, M1b, M2, A1 and A2), we extracted the information of the gas temperature from the corresponding best-model result. Figure~\ref{fig:temperature} shows the behavior of the temperature as a function of the depth of the \hii\, region for each of those models. We estimate the electron density using 9700~K, the median temperature of the range covered by those five gas profiles before reaching the ionization front (where the temperature drop off): from $10^{3.89}$~K to $10^{4.07}$~K. Figure~\ref{fig:ne_oiii} and Figure~\ref{fig:ne_siii} show with a gray area the uncertainties range corresponding to the gas temperature range.

\end{document}